\documentclass[twocolumn,showpacs,aps,prd]{revtex4-1}
\usepackage{graphicx}
\usepackage{float}
\usepackage{epsfig,graphics,subfigure,psfrag,amsmath,amssymb}
\usepackage{lineno}
\usepackage{dcolumn}
\usepackage{bm}
\usepackage{overpic}
\usepackage{xspace}
\usepackage{rotating}
\usepackage{epstopdf}
\usepackage{makecell}
\usepackage{multirow}
\usepackage[colorlinks,linkcolor=blue,anchorcolor=blue,citecolor=blue]{hyperref}

\begin{document}

\title{\bf \boldmath Amplitude analysis and branching-fraction measurement of $D^{+}_{s}\rightarrow K^{0}_{S}K^{-}\pi^{+}\pi^{+}$}

\author{
\begin{small}
\begin{center}
M.~Ablikim$^{1}$, M.~N.~Achasov$^{10,c}$, P.~Adlarson$^{67}$, S. ~Ahmed$^{15}$, M.~Albrecht$^{4}$, R.~Aliberti$^{28}$, A.~Amoroso$^{66A,66C}$, M.~R.~An$^{32}$, Q.~An$^{63,49}$, X.~H.~Bai$^{57}$
% ...
\\
\vspace{0.2cm}
(BESIII Collaboration)\\
\vspace{0.2cm} {\it
$^{1}$ Institute of High Energy Physics, Beijing 100049, People's Republic of China\\
% ...
$^{4}$ Bochum Ruhr-University, D-44780 Bochum, Germany\\
% ...
$^{10}$ G.I. Budker Institute of Nuclear Physics SB RAS (BINP), Novosibirsk 630090, Russia\\
% ...
$^{15}$ Helmholtz Institute Mainz, Johann-Joachim-Becher-Weg 45, D-55099 Mainz, Germany\\
% ...
$^{28}$ Johannes Gutenberg University of Mainz, Johann-Joachim-Becher-Weg 45, D-55099 Mainz, Germany\\
% ...
$^{32}$ Liaoning Normal University, Dalian 116029, People's Republic of China\\
% ...
$^{49}$ State Key Laboratory of Particle Detection and Electronics, Beijing 100049, Hefei 230026, People's Republic of China\\
% ...
$^{57}$ University of Jinan, Jinan 250022, People's Republic of China\\
% ...
$^{63}$ University of Science and Technology of China, Hefei 230026, People's Republic of China\\
% ...
$^{66}$ University of Turin and INFN, (A)University of Turin, I-10125, Turin, Italy; (B)University of Eastern Piedmont, I-15121, Alessandria, Italy; (C)INFN, I-10125, Turin, Italy\\
$^{67}$ Uppsala University, Box 516, SE-75120 Uppsala, Sweden\\
% ...
$^{c}$ Also at the Novosibirsk State University, Novosibirsk, 630090, Russia\\
% ...
}
\end{center}
\vspace{0.4cm}
\end{small}
}

\begin{abstract}
	Using 6.32 fb$^{-1}$ of $e^+e^-$ collision data collected by the BESIII detector at the center-of-mass energies between 4.178 and 4.226~GeV,~an amplitude analysis of the $D^{+}_{s}\rightarrow K^{0}_{S}K^{-}\pi^{+}\pi^{+}$ decays is performed for the first time to determine the intermediate-resonant contributions. The dominant component is the $D_s^+ \to K^*(892)^+\overline{K}^*(892)^0$ decay with a fraction of $(40.6\pm2.9_{\rm stat}\pm4.9_{\rm sys})$\%. Our results of the amplitude analysis are used to obtain a more precise measurement of the branching fraction of the $D^{+}_{s}\rightarrow K^{0}_{S}K^{-}\pi^{+}\pi^{+}$ decay, which is determined to be  $(1.46\pm0.05_{\rm stat}\pm0.05_{\rm sys}$)\%. 
\end{abstract}

\maketitle
\section{Introduction}
The decay $D^{+}_{s}\rightarrow K^{0}_{S}K^{-}\pi^{+}\pi^{+}$ is usually used as a ``tag mode" for  measurements related to the $D_s^+$ meson \cite{a1,a2,a3,a4,a5} due to its large branching fraction and low background contamination. The inclusion of charge-conjugate states is implied throughout the paper.  In 2013 the CLEO Collaboration reported its branching fraction $\mathcal{B}(D^{+}_{s}\rightarrow K^{0}_{S}K^{-}\pi^{+}\pi^{+})$ to be $(1.64\pm0.07\pm0.08)\%$, based on a data sample corresponding to an integrated luminosity of 586~pb$^{-1}$ of $e^{+}e^{-}$ collisions at a center-of-mass energy ($E_{\rm cm}$) of 4.17~GeV \cite{3}. The measurement was limited by the sample size and lack of knowledge of the intermediate processes. In addition, the branching fraction of $D^{+}_{s}\rightarrow K^*(892)^+\overline{K}^*(892)^0$ was determined by the ARGUS Collaboration~\cite{45} more than twenty years ago, who claimed the contribution of $D_s^+ \to K^*(892)^+\overline{K}^*(892)^0$ in the $D^{+}_{s}\rightarrow K^{0}_{S}K^{-}\pi^{+}\pi^{+}$ decays is almost 100\%. The ARGUS measurement suffers from low statistics and large uncertainties in the branching fraction of the reference decay $D^{+}_{s}\rightarrow \phi(1020)\pi^+$. An amplitude analysis of the $D^{+}_{s}\rightarrow K^{0}_{S}K^{-}\pi^{+}\pi^{+}$ decays is necessary to investigate the resonant contributions, and thereby reduce the systematic uncertainties of its branching fraction and for providing input to measurements where amplitude information is essential. 

It is well known that two-body modes  dominate $D_s^+$ decays~\cite{PDG}.~The majority of the observed two-body decays have pseudoscalar-pseudoscalar or pseudoscalar-vector mesons in the final states. Among various kinds of $D_s^+$ decay modes, vector-vector final states are of special interest. The ratios between different orbital angular momenta of the two vector mesons for the dominant quasi-two-body decay $D_{s}^+\rightarrow K^*(892)^+\overline{K}^*(892)^0$ provide valuable information on $CP$ violation with T-violating triple-products~\cite{PhysLettB684}. In addition, several mesons with $J^{P} = 0^-,1^+$ are reported in the mass region between 1.2 and 1.6 GeV/$c^2$ and decay to the $(K\overline{K}\pi)^0$ final state \cite{eta1475_1,eta1475_2,eta1475_3,eta1475_4}.
These are the $\eta(1295)$, $\eta(1405)$, $\eta(1475)$, $f_1(1285)$, $f_1(1420)$ and $f_1(1510)$, although many of these states are not well established.

This paper presents the first amplitude analysis and an improved branching-fraction measurement of the $D^{+}_{s}\rightarrow K^{0}_{S}K^{-}\pi^{+}\pi^{+}$ decay with data samples corresponding to a total integrated luminosity of 6.32 fb$^{-1}$ collected by the BESIII detector at $E_{\rm cm}$ between 4.178 and 4.226 GeV.

\section{DETECTOR AND DATA SETS}
The detailed description of the BESIII detector can be found in Ref.~\cite{Ablikim:2009aa}. It is a magnetic spectrometer located at the Beijing Electron Positron Collider (BEPCII)~\cite{Yu:IPAC2016-TUYA01}. The cylindrical core of the BESIII detector consists of a helium-based multilayer drift chamber (MDC), a plastic scintillator time-of-flight system (TOF), and a CsI(Tl) electromagnetic calorimeter (EMC), which are all enclosed in a superconducting solenoidal magnet providing a 1.0~T magnetic field. The solenoid is supported by an octagonal flux-return yoke with resistive plate counter muon-identifier modules interleaved with steel. The acceptance of charged particles and photons is 93\% over the $4\pi$ solid angle. The charged-particle momenta resolution at 1.0 GeV/$c$ is $0.5\%$, and the specific energy loss ($dE/dx$) resolution is $6\%$ for the electrons from Bhabha scattering. The EMC measures photon energies with a resolution of $2.5\%$ ($5\%$) at $1$~GeV in the barrel (end-cap) region. The time resolution of the TOF barrel part is 68~ps, while that of the end cap part is 110~ps. The end-cap TOF was upgraded in 2015 with multi-gap resistive plate chamber technology, providing a time resolution of 60~ps~\cite{detector1}.

The data samples used in this paper were accumulated in the years 2013, 2016 and 2017 with $E_{\rm cm}$ of 4.226, 4.178 and 4.189$-$4.219 GeV, respectively.
Generic Monte Carlo samples (GMC) that are 40 times larger than the data sets are produced with the GEANT4-based software~\cite{MCsample}. The production of open-charm processes directly via $e^+e^-$ annihilation is modeled with the generator {\sc conexc}~\cite{conExc}, which includes the effects of the beam energy spread and
initial state radiation (ISR). The ISR production of vector charmonium states and the
continuum processes are incorporated in {\sc kkmc}~\cite{KKMC}. The known decay modes are generated using {\sc evtgen}~\cite{EvtGen}, which assumes the branching fractions reported by the Particle Data Group (PDG)~\cite{PDG}. The remaining unknown decays from the charmonium states are generated with {\sc lundcharm}~\cite{LundChar}. The final state radiation from charged tracks are simulated by the {\sc photos} package~\cite{Photos package}. The GMC is used to estimate background and optimize selection criteria.

More than 10 million simulated events are generated with an uniform distribution in the phase space of the $D^{+}_{s}\rightarrow K^{0}_{S}K^{-}\pi^{+}\pi^{+}$ decay to perform the normalization in the amplitude fit. Preliminary parameters of the amplitude model are obtained from an initial fit to the data. A signal Monte Carlo (SMC) sample is generated according to the preliminary parameters and is used to validate the fit performance and to estimate the detector efficiency. A final determination of the fit parameters is obtained by fitting the data using the SMC sample for the normalization. 
\section{EVENT SELECTIONS} \label{tag_selection}
The production of $D^{\pm}_s$ candidates is dominated by the process $e^+e^-\rightarrow D^{*+}_sD^-_s$, where the $D_s^{*+}$ meson decays to either $\gamma D_s^+$ or $\pi^0D_s^+$ with branching fractions of (93.5$\pm$0.7)\% and (5.8$\pm$0.7)\%~\cite{PDG}, respectively. A sample of $D_s^-$ mesons is reconstructed first, with nine $D^{-}_s$ prominent hadronic decay modes, as shown in Table \ref{tag_mode}, and is referred to as the ``single tag (ST)" candidates. The signal decay $D^{+}_{s}\rightarrow K^{0}_{S}K^{-}\pi^{+}\pi^{+}$ is reconstructed by selecting two $\pi^+$, one $K^-$ and one $K^{0}_{S}$ candidates from the unused tracks in each ST event, and is referred to as the sample of ``double tag (DT)" candidates.

All charged tracks reconstructed in the MDC must satisfy $|\cos\theta|<0.93$, where $\theta$ is the polar angle with respect to the direction of the positron beam. Except for $K^0_S$ daughters, they must originate from the interaction point with a distance of closest approach less than 1 cm in the transverse plane and less than 10 cm along the beam direction. The d$E$/d$x$ information in the MDC and the time-of-fight information from the TOF are combined and used for particle identification (PID) by forming confidence levels $CL_{K(\pi)}$ for kaon (pion) hypotheses.  Kaon (pion) candidates are required to satisfy $CL_{K(\pi)}>CL_{\pi(K)}$.

For the photon identification, it is required that each electromagnetic shower starts within 700 ns of the event start time and its energy is greater than 25 (50) MeV in the barrel (end cap) with $|\cos\theta|<0.80$ ($|\cos\theta|\in$[0.86, 0.92]). The $\pi^0$ and $\eta$ candidates are reconstructed via diphoton decays ($\pi^{0}/\eta \to \gamma\gamma$) with the invariant mass of the $\gamma\gamma$ combination $M_{\gamma\gamma}\in$[0.115, 0.150] and [0.50, 0.57] GeV/$c^2$, respectively. The value of $M_{\gamma\gamma}$
is constrained to the $\pi^0$ or $\eta$  nominal mass~\cite{PDG} by a kinematic fit, and the $\chi^2$ of the kinematic fit must be less than 30. We reconstruct the $\eta^{\prime}\rightarrow \pi^+\pi^-\eta$ candidates by requiring $M_{\pi^+\pi^-\eta}\in[0.946, 0.970]$ GeV/$c^2$.

The $K^0_S$ candidates are selected by looping over all pairs of tracks with opposite charges, whose distances to the interaction point along the beam direction are within 20 cm. A primary vertex and a secondary vertex \cite{vertex fit} are reconstructed and the decay length between the two is required to be greater than twice its uncertainty. Since the combinatorial background is low, this requirement is not applied for the $D_s^-\rightarrow K^{0}_{S}K^{-}$ decay. The invariant mass $M_{\pi^+\pi^-}$ is required to be in the region [0.487, 0.511] GeV/$c^2$. To prevent an event being retained by both the $D^{-}_{s}\rightarrow K^{0}_{S}K^{-}$ and $D^{-}_{s}\rightarrow K^{-}\pi^{+}\pi^{-}$ selections, the value of  $M_{\pi^+\pi^-}$ is required to be outside of the mass range [0.487, 0.511]~GeV/$c^2$ for the $D^{-}_{s}\rightarrow K^{-}\pi^{+}\pi^{-}$ decay.

\section{AMPLITUDE ANALYSIS}
\subsection{Selections for Amplitude Analysis}\label{sig_selection}
The tagged $D_s^-$ candidates are constructed from the $\pi^+$, $K^-$, $\eta$, $\eta^\prime$, $K_S^0$ and $\pi^0$ mesons, while the signal $D_s^+$ candidates are reconstructed from the $K^{0}_{S}$, $K^{-}$ and two $\pi^{+}$ mesons. The requirements on the recoiling mass of the $D_s^{+}$, $M_{\rm rec}$, and the mass of the tagged $D_s^-$, $M_{\rm tag}$, are summarized in Table \ref{tag_mode}. The recoiling mass is calculated as follows:
\begin{equation}\small
	M_{\rm rec} = \sqrt{ (E_{\rm cm}- \sqrt{ \vec{\bf p}_{D_s^+}^2 + m^{2}_{D_s^+}} )^2 -  \vec{\bf p}_{D_s^+} ^2}.
\end{equation}
Here, $\vec{\bf p}_{D_s^+}$ is the three momentum of the $D_s^+$ candidate and $m_{D_{s}^+}$ is its nominal mass~\cite{PDG}.
\begin{table}[!hbtp]
\renewcommand\arraystretch{1.25}
 \centering
 \caption{The requirements of $M_{\rm rec}$ for various energies and $M_{\rm tag}$ for individual single-tagged modes. The $K_S^0$, $\pi^0 (\eta)$ and $\eta^{\prime}$ mesons decay to $\pi^+\pi^-$, $\gamma\gamma$ and $\pi^{+}\pi^{-}\eta$ final states, respectively.}
 \vspace{0.05in}
 \begin{tabular}{c|c}
 \hline\hline
 $E_{\rm cm}$ (GeV) &$M_{\rm rec}$ (GeV/$c^2$)\\ \hline
 4.178& [2.050, 2.180]\\ \hline
 4.189& [2.048, 2.190]\\ \hline
 4.199& [2.046, 2.200]\\ \hline
 4.209& [2.044, 2.210]\\ \hline
 4.219& [2.042, 2.220]\\ \hline
 4.226& [2.040, 2.220]\\ \hline\hline
   Tag mode & $M_{\rm tag}$ (GeV/$c^2$)  \\  \hline
 \multicolumn{1}{l|}{$D^{-}_{s}\rightarrow K^{0}_{S}K^{-}$}& [1.948, 1.991]\\ \hline
 \multicolumn{1}{l|}{$D^{-}_{s}\rightarrow K^{+}K^{-}\pi^{-}$} & [1.950, 1.986]  \\ \hline
 \multicolumn{1}{l|}{$D^{-}_{s}\rightarrow K^{0}_{S}K^{-}\pi^{0}$} & [1.946, 1.987]  \\ \hline
 \multicolumn{1}{l|}{$D^{-}_{s}\rightarrow K^{-}\pi^{+}\pi^{-}$} & [1.953, 1.983]  \\ \hline
 \multicolumn{1}{l|}{$D^{-}_{s}\rightarrow \pi^{-}\eta^{\prime}$} & [1.940, 1.996] \\ \hline
 \multicolumn{1}{l|}{$D^{-}_{s}\rightarrow K^{0}_{S}K^{-}\pi^{+}\pi^{-}$} & [1.958, 1.980] \\ \hline
 \multicolumn{1}{l|}{$D^{-}_{s}\rightarrow K^{+}K^{-}\pi^{-}\pi^{0}$} & [1.947, 1.982] \\ \hline
 \multicolumn{1}{l|}{$D^{-}_{s}\rightarrow \pi^{+}\pi^{-}\pi^{-}$} & [1.952, 1.982]  \\ \hline
 \multicolumn{1}{l|}{$D^{-}_{s}\rightarrow \pi^{-}\eta$}  & [1.930, 2.000] \\
 \hline\hline
 \end{tabular}
 \label{tag_mode}
\end{table}

Kinematic fits are performed of the process $e^+e^-\to D_s^{*\pm}D_s^{\mp}\to \gamma D_s^{\pm}D_s^{\mp}$  with the photon assigned to each charm meson in turn, and the $\chi^2$ of the fit being used to decide between the $D_s^{*+}$ and $D_s^{*-}$ hypotheses. The fits include constraints from four-momentum conservation in the $e^+e^-$ system, and also constrain the invariant masses of $K_S^0$, $D_s^{*\pm}$ and tag-side $D_s^{\pm}$ candidates to their nominal masses~\cite{PDG}.
In order to ensure that all candidates fall within the kinematic boundary of the phase space, we perform a further kinematic fit in which the signal $D_s^\pm$ mass is constrained to  its nominal value, and the updated four-momenta are used for the amplitude analysis.

To suppress the background where the $\pi^-$ from the signal decay
and the $\pi^+$ from the tag modes are exchanged, which fakes the signal and the
same tag mode but with opposite charges,
we perform kinematic fits with $D_s^+$ and $D_s^-$ mass constraints for the two cases, and select the one with the smaller $\chi^2$.
To reduce the background coming from
$D^0 \to K^- \pi^+ \pi^+ \pi^-$ versus $\overline{D}^0 \to K_S^0 K^+ K^- (K_S^0 \pi^+\pi^-)$,
by exchanging $\pi^-$ from $D^0$ and $K_S^0$ from $\overline{D}{}^0$, faking the signal
$D_s^+ \to K_S^0 K^- \pi^+ \pi^+$ and the tag mode $D_s^- \to K^+K^-\pi^- (\pi^+\pi^-\pi^-)$,
we reject events satisfying:
$|M_{D^0}^\prime      -M_{D^0}^{\rm PDG}|<15\mbox{\,MeV}/c^2$,
$|M_{\overline{D}^0}^\prime-M_{D^0}^{\rm PDG}|<15\mbox{\,MeV}/c^2$ and
$|M_{D^0}^\prime-M_{\overline{D}^0}^\prime|<|M_{D_s^+}^\prime-M_{D_s^-}^\prime|$.
Here, $M_{D^0}^{\rm PDG}$ is the nominal $D^0$ mass~\cite{PDG},
$M_{D^0}^\prime$, $M_{\overline{D}^0}^\prime$, $M_{D_s^+}^\prime$ and $M_{D_s^-}^\prime$
are the invariant masses of the $D^0 \to K^-\pi^+\pi^+\pi^-$, $\overline{D}^0 \to K_S^0 K^+K^-(K_S^0\pi^+\pi^-)$,
$D_s^+ \to K_S^0 K^- \pi^+ \pi^+$,  and $D_s^- \to K^+K^-\pi^- (\pi^+\pi^-\pi^-)$ candidates, respectively.

\begin{figure*}[htb]
 \centering
 \begin{overpic}[width=0.30\textwidth,height=0.22\textwidth]{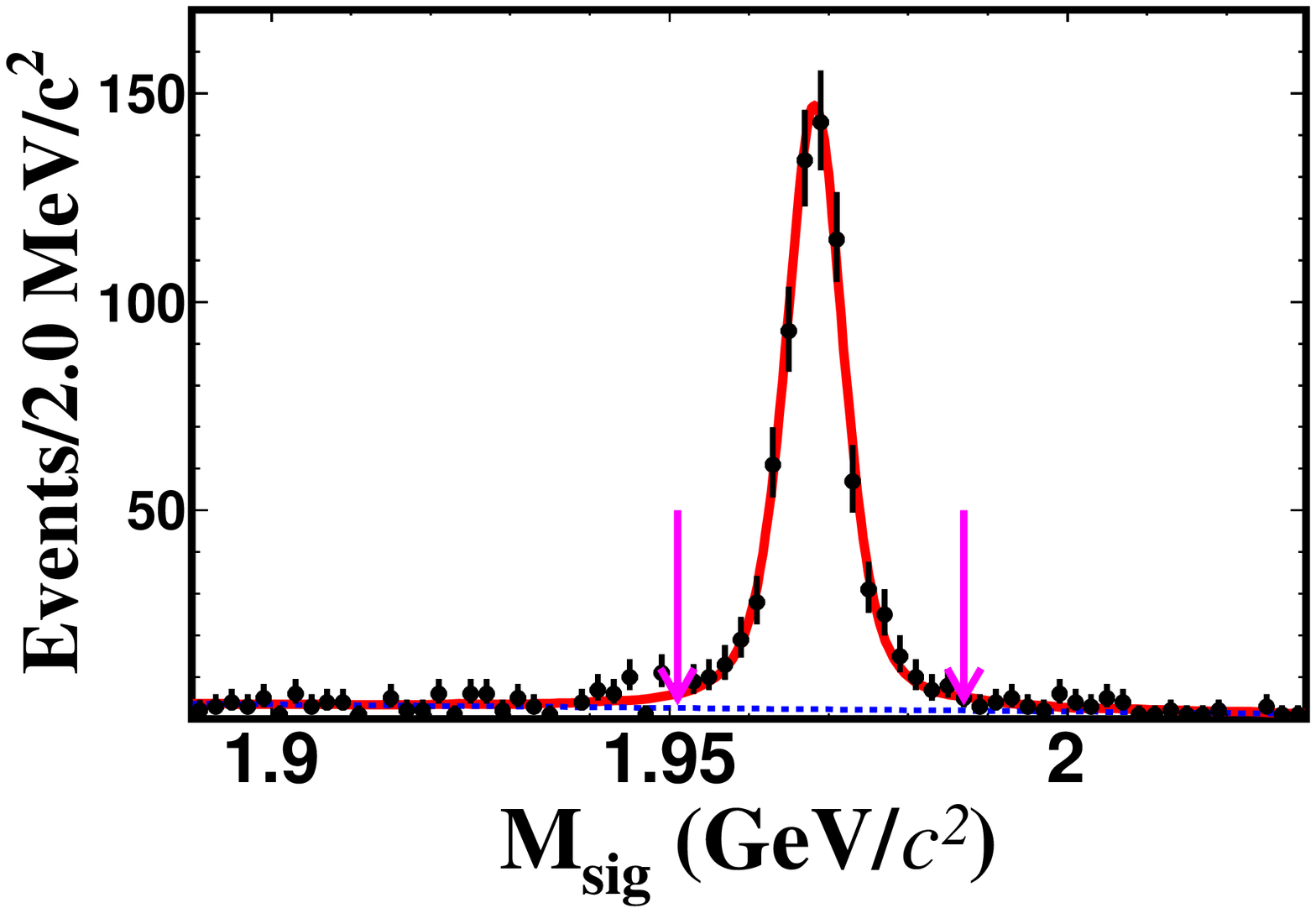}
 \put(80,60){$(a)$}
 \end{overpic}
 \begin{overpic}[width=0.30\textwidth,height=0.22\textwidth]{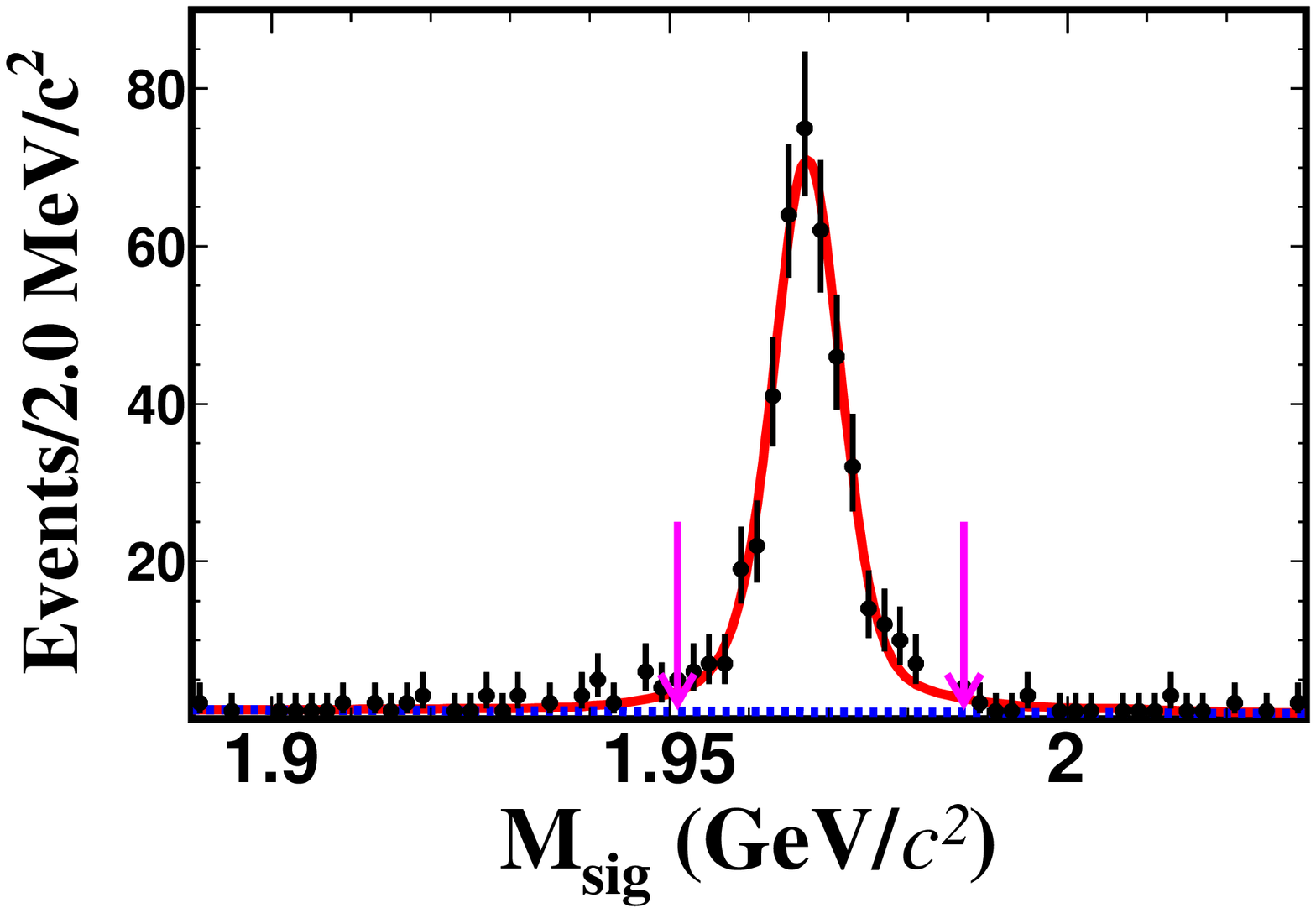}
 \put(80,60){$(b)$}
 \end{overpic}
 \begin{overpic}[width=0.30\textwidth,height=0.22\textwidth]{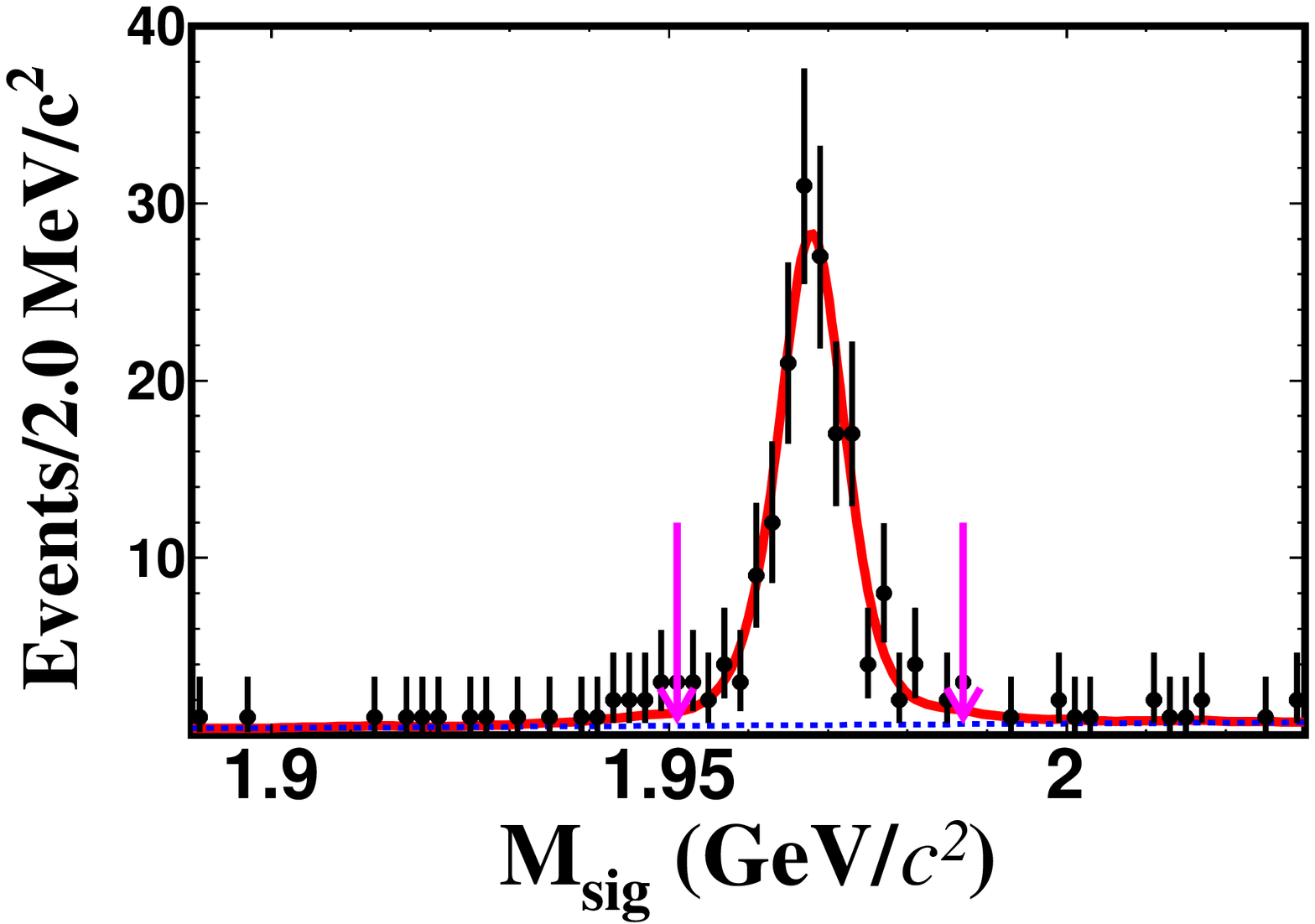}
 \put(80,60){$(c)$}
 \end{overpic}
 \caption{Fits to the $M_{\rm sig}$ distributions of accepted candidates from the data samples taken at (a) $E_{\rm cm}$ = 4.178, (b) 4.189$-$4.219 and (c) 4.226 GeV, respectively. The points with error bars are data. The red solid curves are the fit results. The blue dotted curves are the fitted background shapes. The pair of pink arrows indicate the chosen signal region.}
\label{purity}
\end{figure*}

Figure~\ref{purity} shows the invariant mass distributions of the signal $D_s^+$, $M_{\rm sig}$, in data and the fit results. The signal distribution is modeled with the simulated shape convolved with a Gaussian function and the background is described by a first-order Chebychev polynomial. The fitted yields for the signal are 744$\pm$28, 415$\pm$21 and 159$\pm$13 in the invariant mass range [1.951, 1.987] GeV/$c^2$, with purities of (94.7$\pm$0.5)\%, (96.2$\pm$0.7)\% and (93.9$\pm$1.2)\% for the data samples taken at $E_{cm} = 4.178, 4.189-4.219$ and 4.226 GeV, respectively. The candidates falling in the $D_s^+$ mass region are retained for the amplitude analysis. We compare the background yield and various distributions of the events outside the signal region between data and GMC. The yield and distributions are found to be consistent within the statistical uncertainties. The background events in the signal region from GMC are used to estimate the background contributions in data.
\subsection{Likelihood Function}\label{sec:PDF_likelihood_fit}
An unbinned-maximum-likelihood method is applied to determine resonant contributions in the $D^{+}_{s}\rightarrow K^{0}_{S}K^{-}\pi^{+}\pi^{+}$ decays. The likelihood function is constructed with a probability density function (PDF) of the momenta of the four daughter particles. 
The amplitude of the $n^{\rm th}$ intermediate state ($A_n$) is
\begin{equation}
A_n = P^1_nP^2_nS_nF^1_nF^2_nF^{3}_n,
\end{equation}
where the indices 1, 2 and 3 correspond to the two subsequent intermediate resonances and the $D_s^+$ meson. $S$ is the spin factor constructed with the covariant tensor formalism~\cite{Zou2003}, $F$ is the Blatt-Weisskopf barrier factor and $P$ is the propagator of the intermediate resonance. The total amplitude $M$ is a coherent sum of the amplitudes of intermediate processes,
\begin{equation}
M = \sum{c_nA_n},
\label{rho_phi}
\end{equation}
where $c_n = \rho_ne^{i\phi_n}$ are
complex coefficients to be determined from the fit to data. 

The signal PDF $f_S(p_j)$ is given by
\begin{equation}
f_S(p_j) = \frac{\epsilon(p_j)|M(p_j)|^2R_4(p_j)}{\int \epsilon(p_j)|M(p_j)|^2R_4(p_j)dp_j},
\end{equation}
where $\epsilon(p_j)$ is the detection efficiency parameterized in terms of the final four-momenta $p_j$ and $j$ refers to the different particles in the final
states. $R_4(p_j)$ is the standard element of the four-body phase space. 

The normalization is determined from the simulated events,
\begin{equation}\small
\int \epsilon(p_j)|M(p_j)|^2R_4(p_j)dp_j \approx \frac{1}{N_{\rm sim}}\sum^{N_{\rm sim}}_{k_{\rm sim}}\frac{|M(p^{k_{\rm sim}}_j)|^2}{|M^{\rm gen}(p^{k_{\rm sim}}_j)|^2},
\label{MC intergration}
\end{equation}
where $k_{\rm sim}$ runs from 1 to $N_{\rm sim}$, the total number of simulated events. $M^{\rm gen}(p_j)$ is the PDF used to generate the simulated samples.

The normalization takes into account the difference in detector
efficiencies for PID and tracking between data and simulation by assigning a weight to each simulated event
\begin{equation}
\label{pid_fun}
\gamma_{\epsilon}(p) = \prod_{i} \frac{\epsilon_{i,\rm data}(p_j)}{\epsilon_{i,\rm sim}(p_j)},
\end{equation}
where $i$ denotes the four daughter particles. The normalization is then given by
\begin{equation}\footnotesize
  \int \epsilon(p_j)|M(p_j)|^2R_4(p_j)dp_j \approx \frac{1}{N_{\rm sim}}\sum^{N_{\rm sim}}_{k_{\rm sim}}\frac{\gamma_{\epsilon}(p_j^{k_{\rm sim}})|M(p^{k_{\rm sim}}_j)|^2}{|M^{\rm gen}(p^{k_{\rm sim}}_j)|^2}.
\end{equation}

The total PDF $f_T(p_j)$ is
\begin{equation}
\begin{aligned}
f_T(p_j)& = w\frac{\epsilon(p_j)|M(p_j)|^{2}R_{4}(p_j)}{\int \epsilon(p_j)|M(p_j)|^{2}R_{4}(p_j)dp_j}\\
&+ (1-w)\frac{B(p_j)R_{4}(p_j)}{\int B(p_j)R_{4}(p_j)dp_j},
\end{aligned}
\end{equation}
where $\mathit{w}$ is the purity of the signal described by a constant parameter in the fit. We factorize out $\epsilon(p_j)$ from $f_T(p_j)$ as $\epsilon(p_j)$ is independent of the fit variables. Its contribution enters into the normalization and the background PDF.
As a consequence, the combined PDF becomes
\begin{equation}
\begin{aligned}
f_T(p_j)& = \epsilon(p_j)R_4(p_j)[w\frac{|M(p_j)|^2}{\int \epsilon(p_j)|M(p_j)|^2R_4(p_j)dp_j} \\
&+ (1-w)\frac{B_{\epsilon}(p_j)}{\int \epsilon(p_j)B_{\epsilon}(p_j)R_4(p_j)dp_j}],
\label{bkg_frac}
\end{aligned}
\end{equation}
where $B_{\epsilon}(p_j) = B(p_j)/\epsilon$ and the background PDF $B(p_j)$ is parameterized using RooNDKeysPdf~\cite{RooNDKeysPdf}. The normalization in the denominator of the background term is calculated as
\begin{small}
\begin{equation}
  \int \epsilon(p_j)B_{\epsilon}(p_j)R_4(p_j)dp_j \approx \frac{1}{N_{\rm sim}}\sum^{N_{\rm sim}}_{k_{\rm sim}}\frac{B_{\epsilon}(p_j^{k_{\rm sim}})}{|M^{\rm gen}(p^{k_{\rm sim}}_j)|^2}.
\end{equation}
\end{small}

Finally the log-likelihood is written as
\begin{equation}
\begin{aligned}
{\rm ln}\mathcal{L} &= {\rm ln}[w\frac{|M(p_j)|^2}{\int \epsilon(p_j)|M(p_j)|^2R_4(p_j)dp_j} \\
&+ (1-w)\frac{B_{\epsilon}(p_j)}{\int \epsilon(p_j)B_{\epsilon}(p_j)R_4(p_j)dp_j}],
\end{aligned}
\end{equation}
and data samples collected at different $E_{cm}$ are fitted simultaneously. 
\subsection{Spin Factors}\label{ch:spin}
For the process $a \rightarrow bc$, the four momenta of the particles $a$, $b$ and $c$ are denoted as $p_a$, $p_b$ and $p_c$, respectively.
The spin projection operators~\cite{Zou2003} are defined as
\begin{eqnarray}
\begin{aligned}
P^{(1)}_{\mu\mu^{\prime}}(a) &= -g_{\mu\mu^{\prime}}+\frac{p_{a,\mu}p_{a,\mu^{\prime}}}{p^2_a}\,,\\
P^{(2)}_{\mu\nu\mu^{\prime}\nu^{\prime}}(a) &= \frac{1}{2}(P^{(1)}_{\mu\mu^{\prime}}(a)P^{(1)}_{\nu\nu^{\prime}}(a)+P^{(1)}_{\mu\nu^{\prime}}(a)P^{(1)}_{\nu\mu^{\prime}}(a))\\
&-\frac{1}{3}P^{(1)}_{\mu\nu}(a)P^{(1)}_{\mu^{\prime}\nu^{\prime}}(a)\,.\end{aligned}
\end{eqnarray}
The pure orbital angular-momentum covariant tensors are given by
\begin{eqnarray}
\begin{aligned}
\tilde{t}^{(1)}_{\mu}(a)&= -P^{(1)}_{\mu\mu^{\prime}}(a)r_a^{\mu^{\prime}}\,,\\
\tilde{t}^{(2)}_{\mu\nu}(a)&= P^{(2)}_{\mu\nu\mu^{\prime}\nu^{\prime}}(a)r_a^{\mu^{\prime}}r_a^{\nu^{\prime}}\,,
\end{aligned}
\end{eqnarray}
where $r_a\,=\,p_b - p_c$. The spin factors $S(p)$ used in this paper are constructed from the spin projection operators and pure orbital angular-momentum covariant tensors and are listed in Table~\ref{table:spin_factors}.
\begin{table}[!hbtp]
\renewcommand\arraystretch{1.45}
 \begin{center}
 \caption{The spin factor $S(p)$ for each decay chain. All operators, i.e.~$\tilde{t}$, have the same definitions as in Ref.~\cite{Zou2003}.
  Scalar, pseudo-scalar, vector and axial-vector states are denoted
  by $S$, $P$, $V$ and $A$, respectively. The $[S]$, $[P]$ and $[D]$ denote the orbital angular-momentum quantum numbers $L$ = 0, 1 and 2, respectively.}\label{table:spin_factors}
\begin{tabular}{ll}
 \hline
\hline
 Decay chain& $S(p)$ \\
 \hline
$D_s^+[S] \rightarrow V_1V_2$ & $\tilde{t}^{(1)\mu}(V_1) \; \tilde{t}^{(1)}_\mu(V_2)$ \\ 
$D_s^+[P] \rightarrow V_1V_2$ & $\epsilon_{\mu\nu\lambda\sigma}p^\mu(D_s^+) \; \tilde{T}^{(1)\nu}(D_s^+) \;$ \\ 
                                                    & $\times  \tilde{t}^{(1)\lambda}(V_1) \; \tilde{t}^{(1)\sigma}(V_2) $ \\
$D_s^+[D] \rightarrow V_1V_2$ & $\tilde{T}^{(2)\mu\nu}(D_s^+) \; \tilde{t}^{(1)}_\mu(V_1) \; \tilde{t}^{(1)}_\nu(V_2)$\\
$D_s^+ \rightarrow AP_1, A[S] \rightarrow VP_2$ & $\tilde{T}^{(1)\mu}(D_s^+) \; P^{(1)}_{\mu\nu}(A) \; \tilde{t}^{(1)\nu}{(V)}$ \\
$D_s^+ \rightarrow AP_1, A \rightarrow SP_2$ &$\tilde{T}^{(1)\mu}(D_s^+)\tilde{t}^{(1)}_\mu(A)$ \\
$D_s^+ \rightarrow VS$ & $\tilde{T}^{(1)\mu}(D_s^+)\;\tilde{t}^{(1)}_\mu(V)$ \\
$D_s^+ \rightarrow PP_1, P\rightarrow VP_2 $ & ${p^\mu }({P_2})\tilde{t}_\mu ^{(1)}(V)$ \\
$D_s^+ \rightarrow PP_1, P\rightarrow SP_2 $ & 1 \\
\hline
\hline
\end{tabular}
 \end{center}
\end{table}
\subsection{Blatt-Weisskopf Barrier Factors}
For the process $a \rightarrow bc$, the Blatt-Weisskopf barrier factor $F_L(p_j)$ is parameterized as a function of the
angular momentum $L$ and the momentum $q$ of the daughter $b$ or $c$ in the rest system of $a$,
\begin{eqnarray}
\begin{aligned}
F_L(q) = z^L X_L(q),
\end{aligned}
\end{eqnarray}
where $z\,=\,qR$. $R$ is the effective radius of the barrier, which is fixed to 3.0 GeV$^{-1} \times \hbar c$ for the intermediate resonances and 5.0 GeV$^{-1} \times \hbar c$ for the $D_s^+$ meson \cite{RD_value}. 
The momentum-transfer squared is
\begin{eqnarray}
\begin{aligned}
q^2 = \frac{(s_a+s_b-s_c)^2}{4s_a}-s_b\,,
\end{aligned}
\end{eqnarray}
where $s_{a, b, c}$ are the invariant-mass squared of particles $a, b, c$, respectively.
The $X_L(q)$ factors are given by
\begin{eqnarray}
\begin{aligned}
 X_{L=0}(q)&=1,\\
 X_{L=1}(q)&=\sqrt{\frac{2}{z^2+1}},\\
 X_{L=2}(q)&=\sqrt{\frac{13}{z^4+3z^2+9}}\,.
\end{aligned}
\end{eqnarray}

\subsection{Propagators}
The propagators for the resonances $K^*(892)^+$, $\overline{K}^*(892)^0$, $\eta(1295)$, $\eta(1405)$, $\eta(1475)$, $f_1(1285)$, $f_1(1420)$ and $f_1(1510)$ are modeled by the relativistic Breit-Wigner function,
       which is given by
\begin{eqnarray}
\begin{aligned}
P(m)&=\frac{1}{(m^2_0-s_a)-im_0\Gamma(m)},\\
\Gamma(m)&=\Gamma_0\left(\frac{q}{q_0}\right)^{2L+1}\left(\frac{m_0}{m}\right)\left(\frac{X_L(q)}{X_L(q_0)}\right)^2\,,
\end{aligned}
\end{eqnarray}
where $m_0$ and $\Gamma(m)$ are the mass and width of the intermediate resonance, and $q_0$ is the value of $q$ when $s_a = m^2_0$.

The $a_0(980)$ contribution is parameterized as the Flatt\'{e} formula
\begin{small}
\begin{equation}
P_{a_0(980)} = \frac{1}{M^2-s_a-i(g_{\eta\pi}\rho_{\eta\pi}(s_a)+g_{K\overline K}\rho_{K\overline K}(s_a))},
\end{equation}
\end{small}where $\rho_{\eta\pi}(s_a)$ and $\rho_{K\overline K}(s_a)$ are the Lorentz-invariant phase-space factors defined as $2q/\sqrt {s_a}$, and the
coupling constants $g^{2}_{\eta\pi} = 0.341\pm0.004$ GeV$^2$/$c^4$ and $g^2_{K\overline K} = (0.892\pm0.022)g^2_{\eta\pi}$~\cite{PhysRevD.78.032002}.

We use the same parameterization to describe the $K\pi$ $S$-wave as Ref.~\cite{PRD112012}, which is extracted
from scattering data~\cite{ASTON1988493}. The model is built with a Breit-Wigner shape for the $K^{*}(1430)^0$ and an
effective range parameterization for the non-resonant component,
\begin{eqnarray}\label{eq:kpi_swave_formfactor}
\begin{aligned}
 A(m) = F\sin\delta_F e^{i\delta_F}+R\sin\delta_R e^{i\delta_R}e^{i2\delta_F}\,,
\end{aligned}
\end{eqnarray}
with
\begin{eqnarray}
\begin{aligned}
 \delta_F&=\phi_F+\cot^{-1}\left[\frac{1}{aq}+\frac{rq}{2}\right]\nonumber,\\
 \delta_R&=\phi_R+\tan^{-1}\left[\frac{M\Gamma(m_{K\pi})}{M^2-m^2_{K\pi}}\right]\,,\nonumber
\end{aligned}
\end{eqnarray}
where $a$ and $r$ are the scattering length and effective interaction length, respectively.
The parameters $F(\phi_F)$ and $R(\phi_R)$ are the magnitudes (phases) for the non-resonant term
and the resonant contribution, respectively.
The parameters $M$, $F$, $\phi_F$,
$R$, $\phi_R$, $a$ and $r$ are fixed to the results of the $D^0\rightarrow K^0_S\pi^+\pi^-$ analysis
by the $B${\scriptsize $\mathit A$}$B${\scriptsize $\mathit AR$} and Belle Collaborations~\cite{PRD112012}.
\subsection{Fit Fraction }\label{sec:fit_fraction}
The fit fraction (FF) for a quasi-two-body contribution is independent of any normalization and phase conventions in the amplitude formalism, and hence provides a more useful measure of amplitude strengths than the magnitudes of each contribution. The definition of the FF for the $n^{\text{th}}$ contribution is
\begin{equation}\footnotesize
 {\rm FF}_n = \frac{\int\vert c_nA_n(p)\vert^2R_4(p)dp}{\int\vert\sum\limits_{k} c_kA_k(p)\vert^2R_4(p)dp}
    \approx \frac{\sum\limits_{l=1}^{N_{g,{\rm ph}}}\vert c_nA_n(p^l)\vert^2}{\sum\limits_{l=1}^{N_{g,{\rm ph}}}\vert\sum\limits_k c_kA_k(p^l)\vert^2},
\end{equation}
where the integration is approximated by the sum of the simulated events generated flatly over the phase space and without any efficiency effects included. 

To estimate the statistical uncertainties on the FFs, the calculation is repeated  by randomly varying the fit parameters according to the error matrix. The resulting distribution of each FF is fitted with a Gaussian function, whose width gives the corresponding statistical uncertainty.
\subsection{Fit Results}\label{fit results}
\begin{figure*}[htb]
 \centering
 \begin{overpic}[width=0.30\textwidth,height=0.22\textwidth]{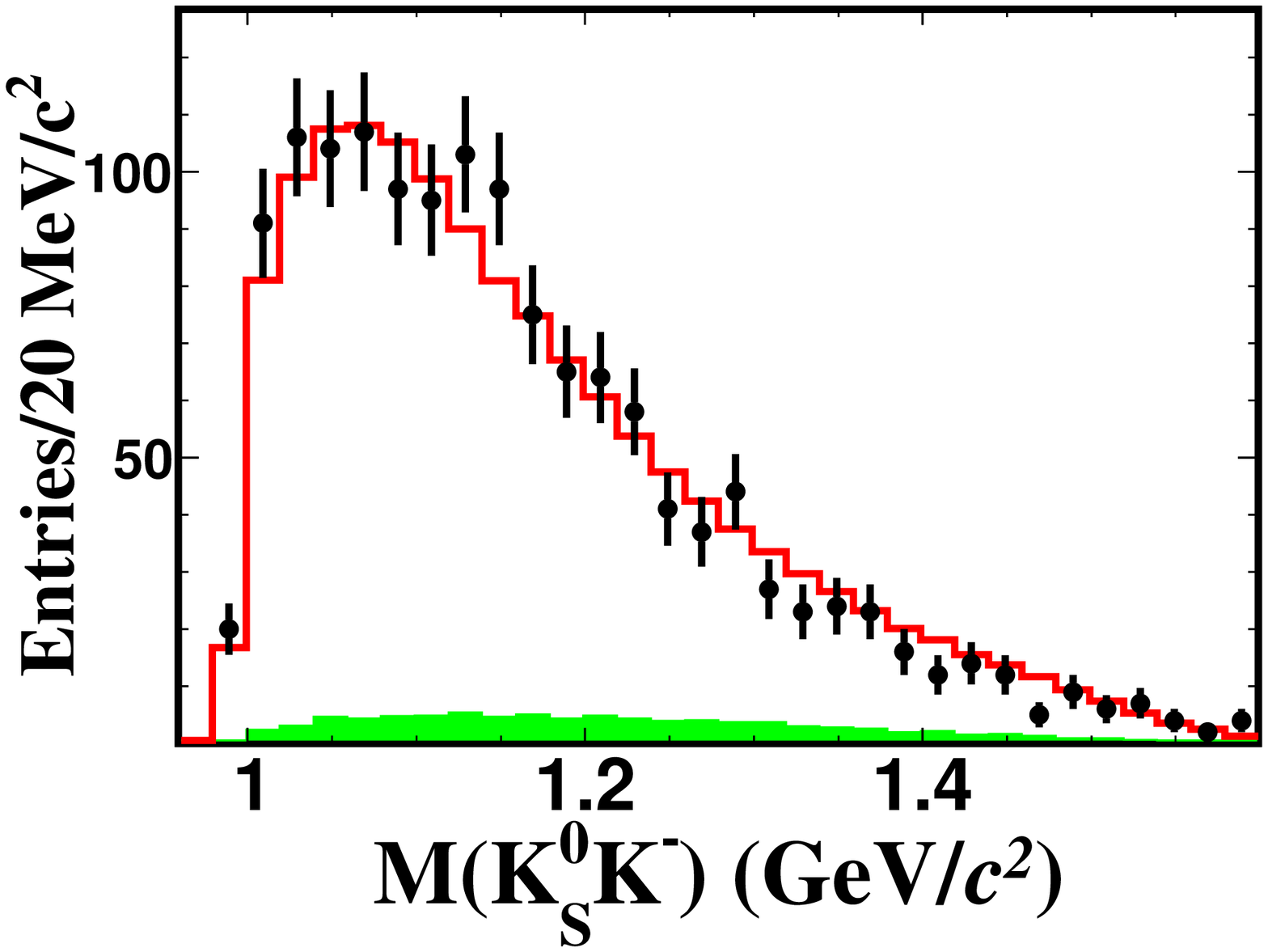}
 \put(85,60){$(a)$}
 \end{overpic}
 \begin{overpic}[width=0.30\textwidth,height=0.22\textwidth]{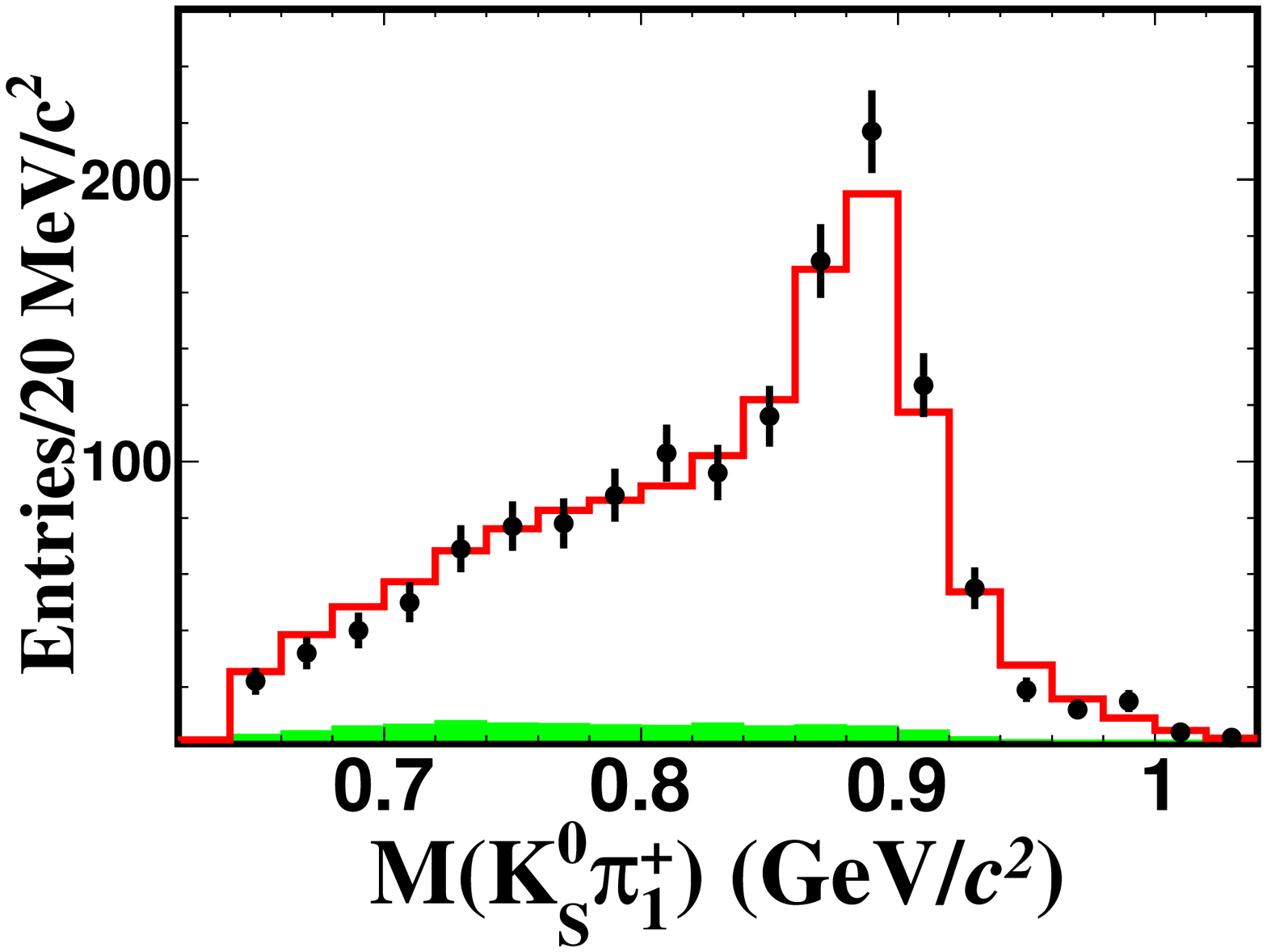}
 \put(85,60){$(b)$}
 \end{overpic}
 \begin{overpic}[width=0.30\textwidth,height=0.22\textwidth]{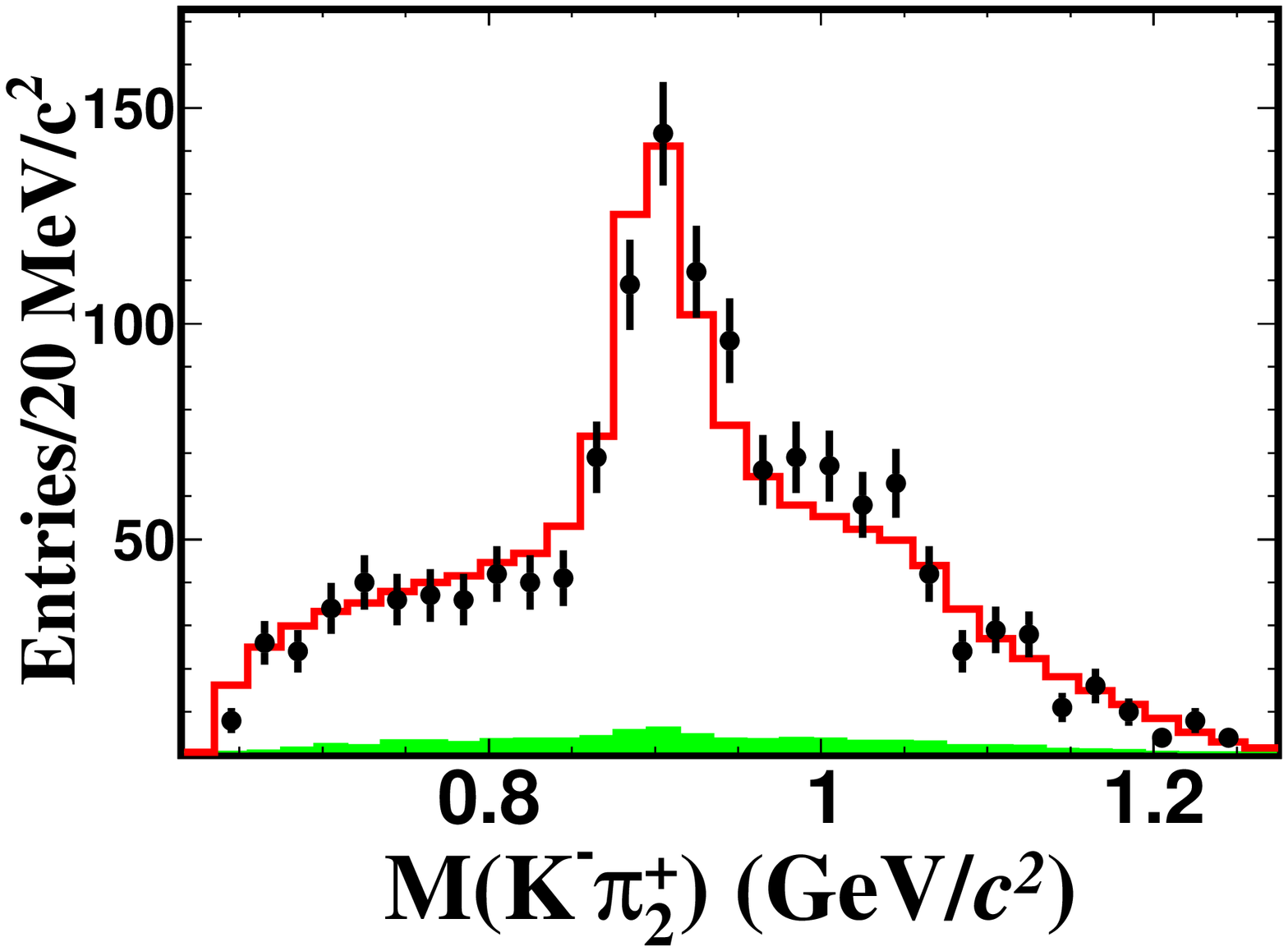}
 \put(85,60){$(c)$}
 \end{overpic}
 \begin{overpic}[width=0.30\textwidth,height=0.22\textwidth]{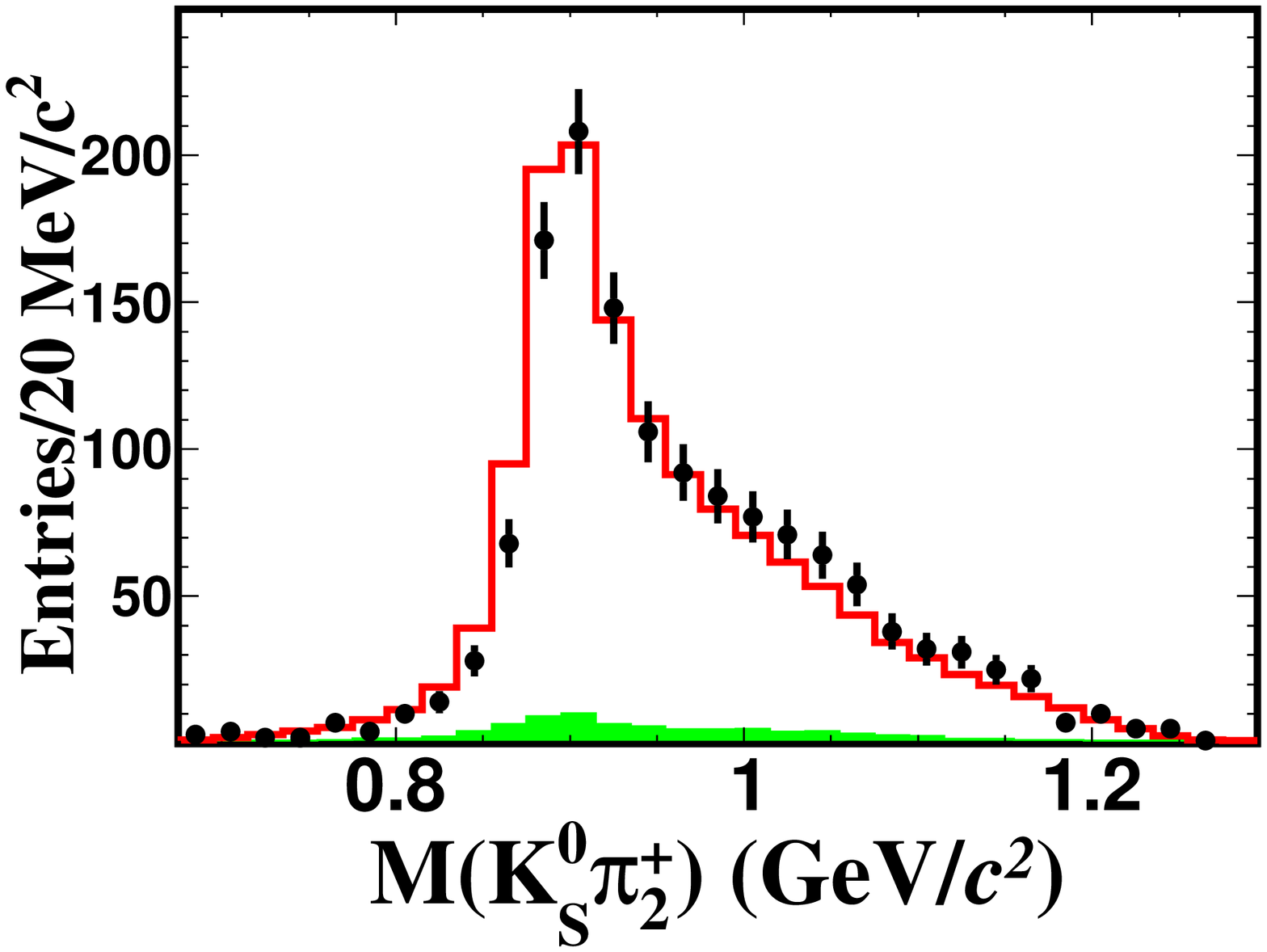}
 \put(85,60){$(d)$}
 \end{overpic}
 \begin{overpic}[width=0.30\textwidth,height=0.22\textwidth]{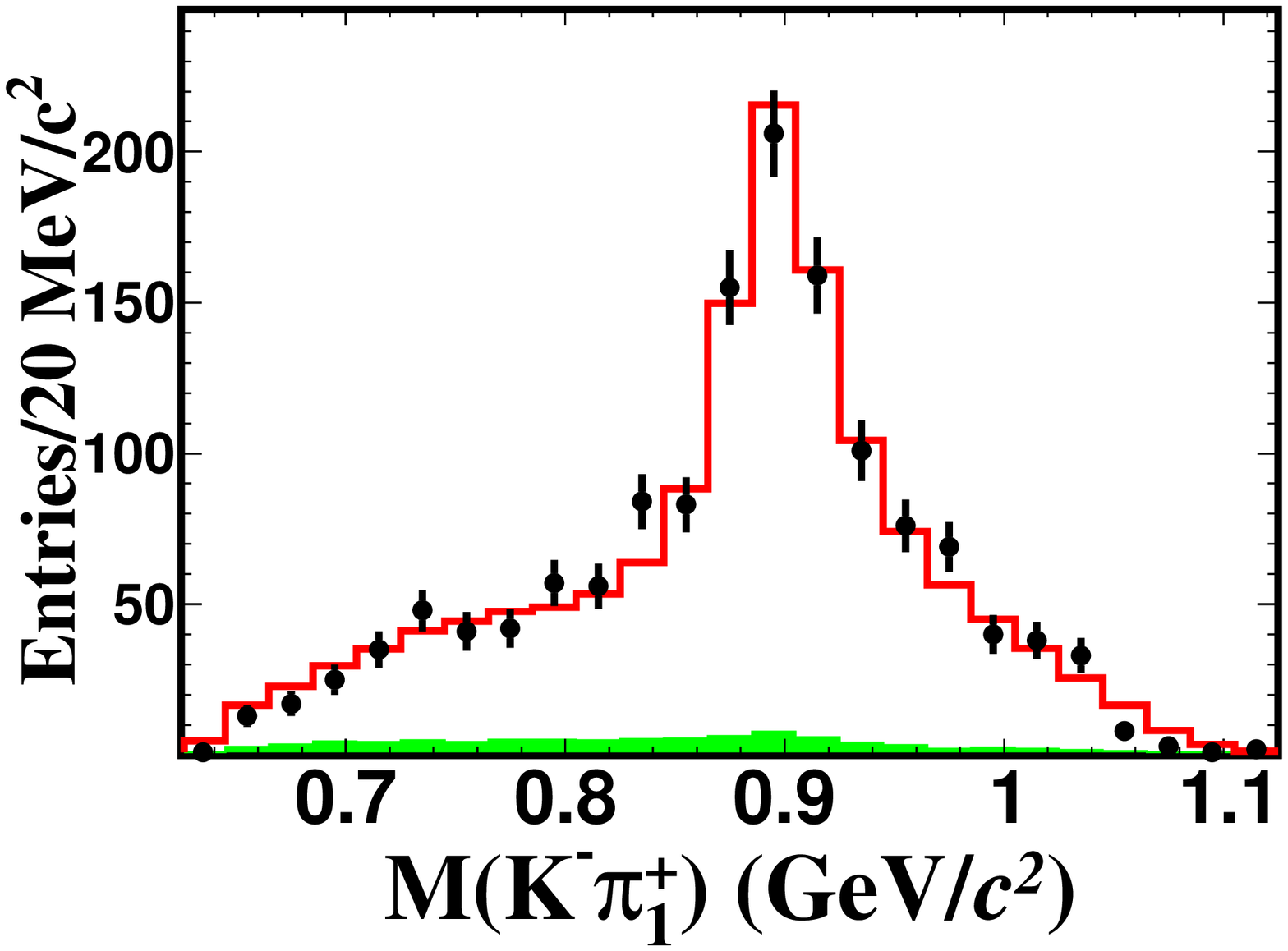}
 \put(85,60){$(e)$}
 \end{overpic}
 \begin{overpic}[width=0.30\textwidth,height=0.22\textwidth]{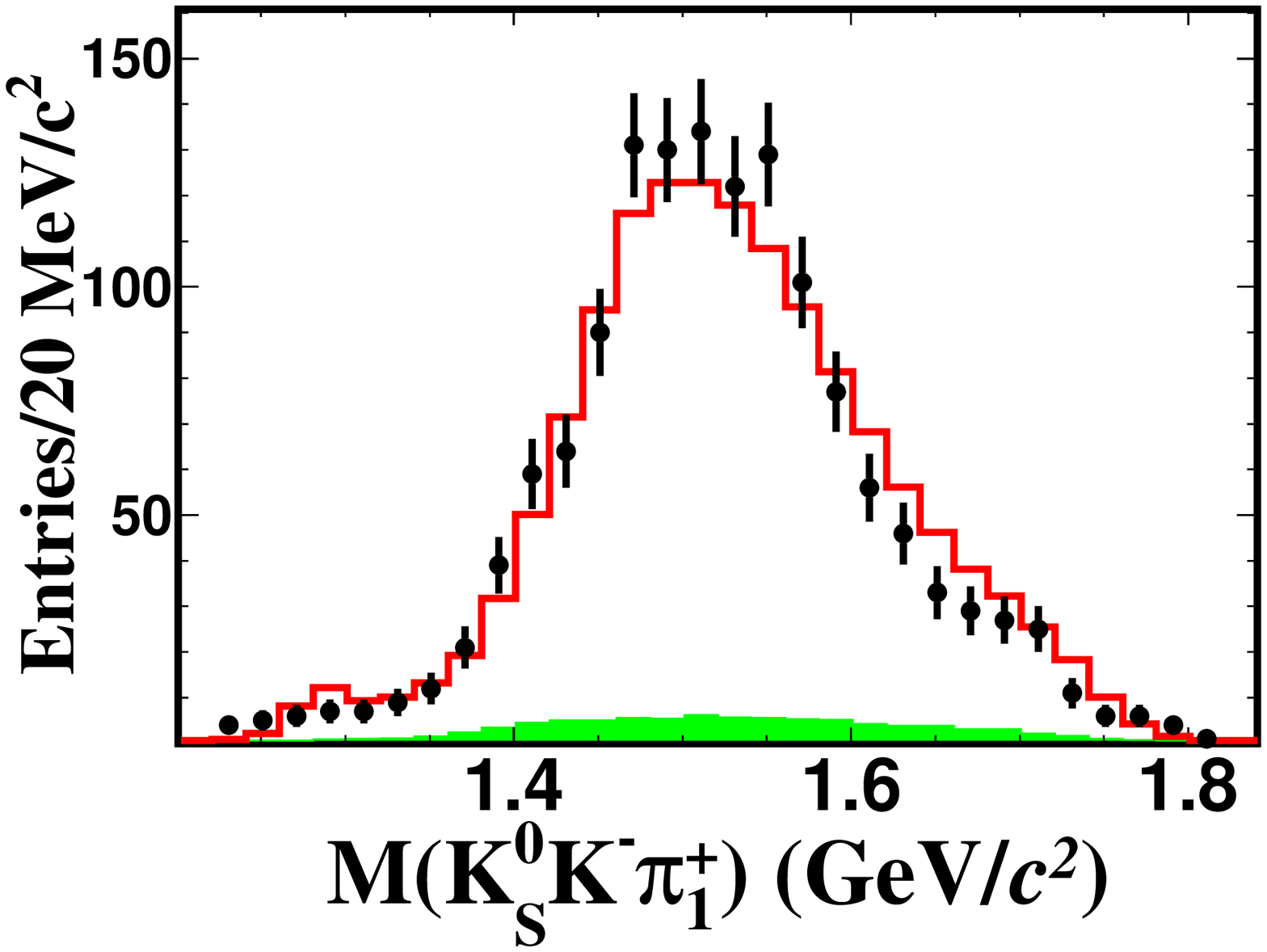}
 \put(85,60){$(f)$}
 \end{overpic}
 \begin{overpic}[width=0.30\textwidth,height=0.22\textwidth]{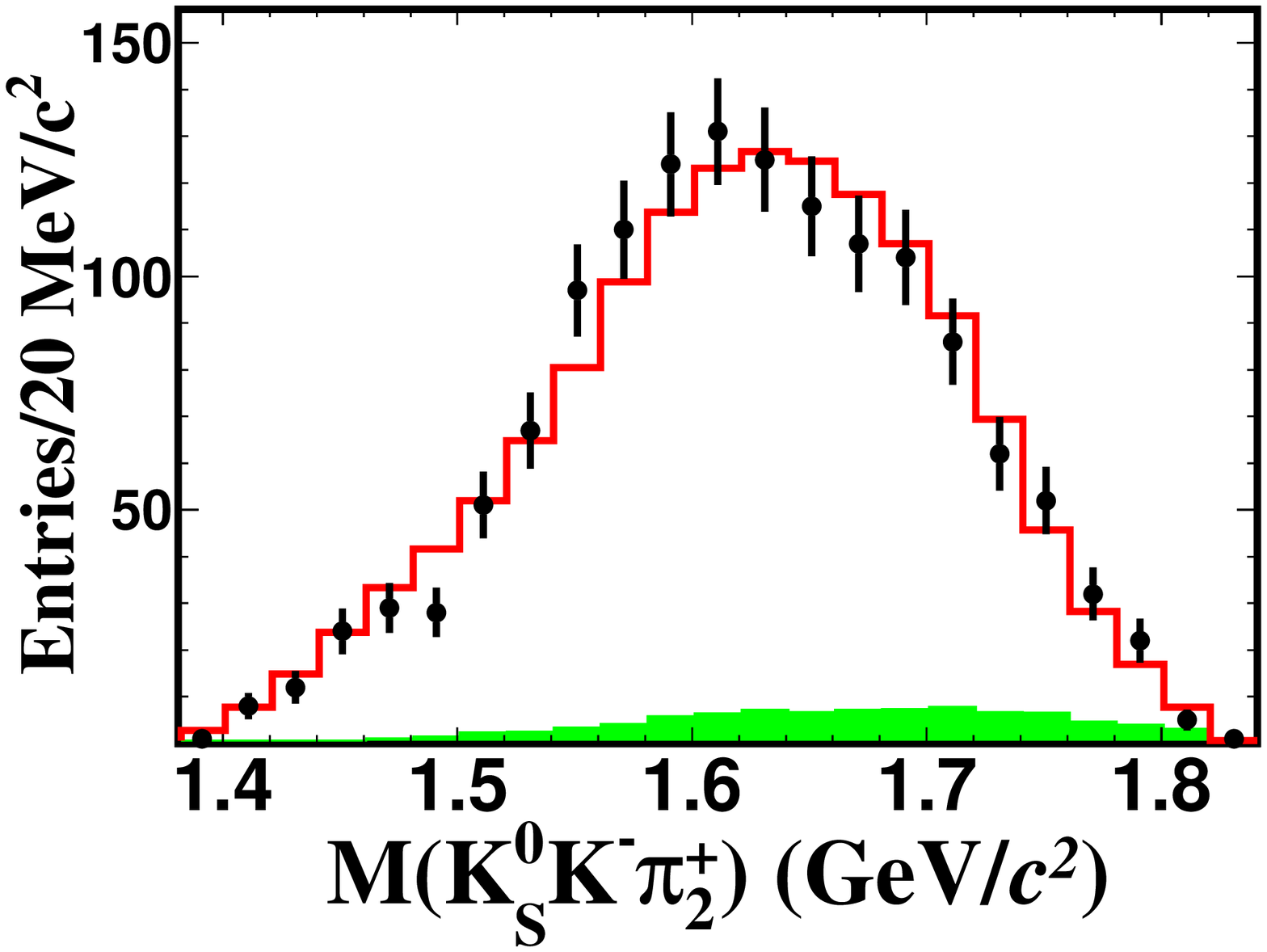}
 \put(85,60){$(g)$}
 \end{overpic}
 \begin{overpic}[width=0.30\textwidth,height=0.22\textwidth]{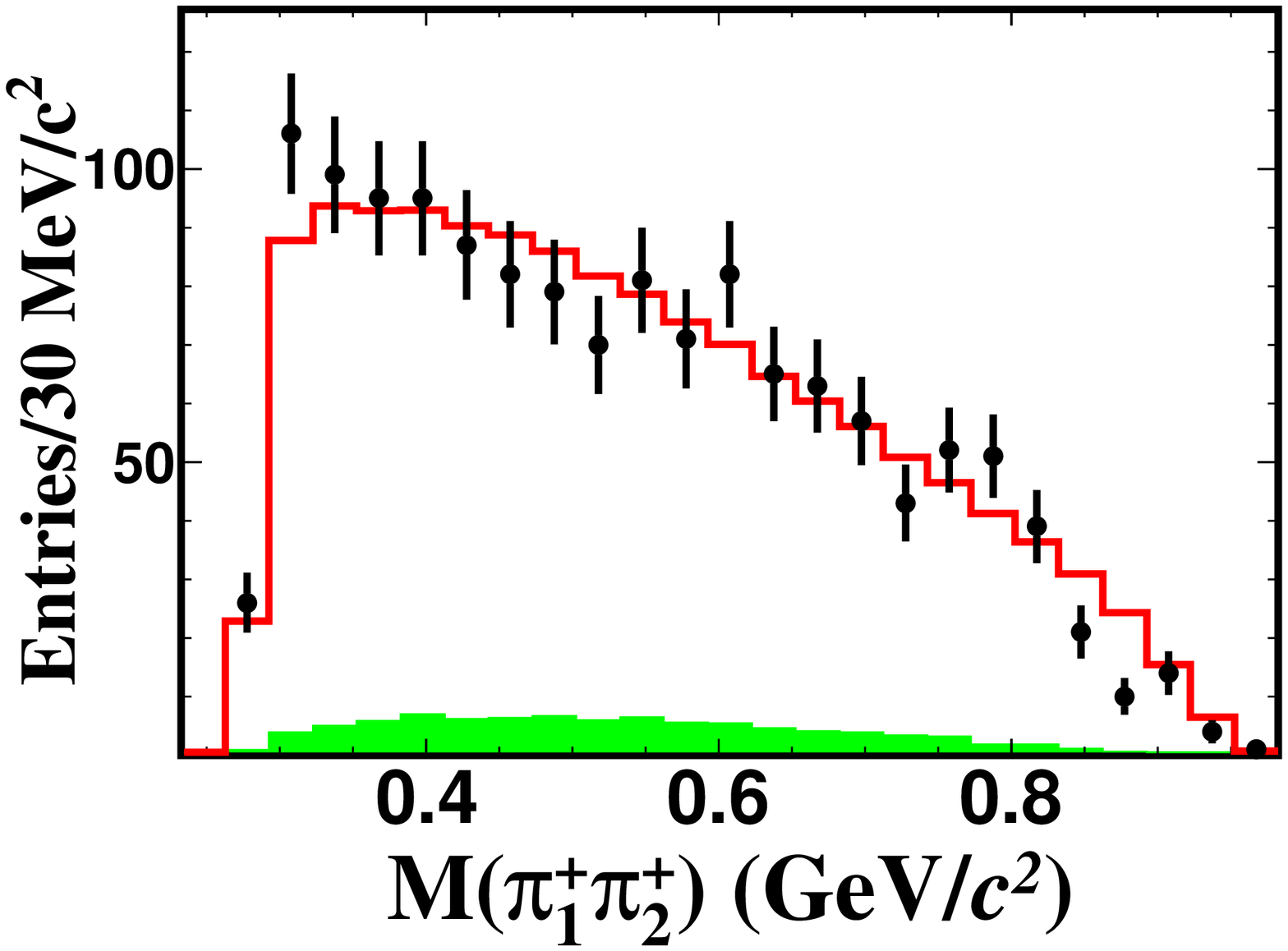}
 \put(85,60){$(h)$}
 \end{overpic}
 \begin{overpic}[width=0.30\textwidth,height=0.22\textwidth]{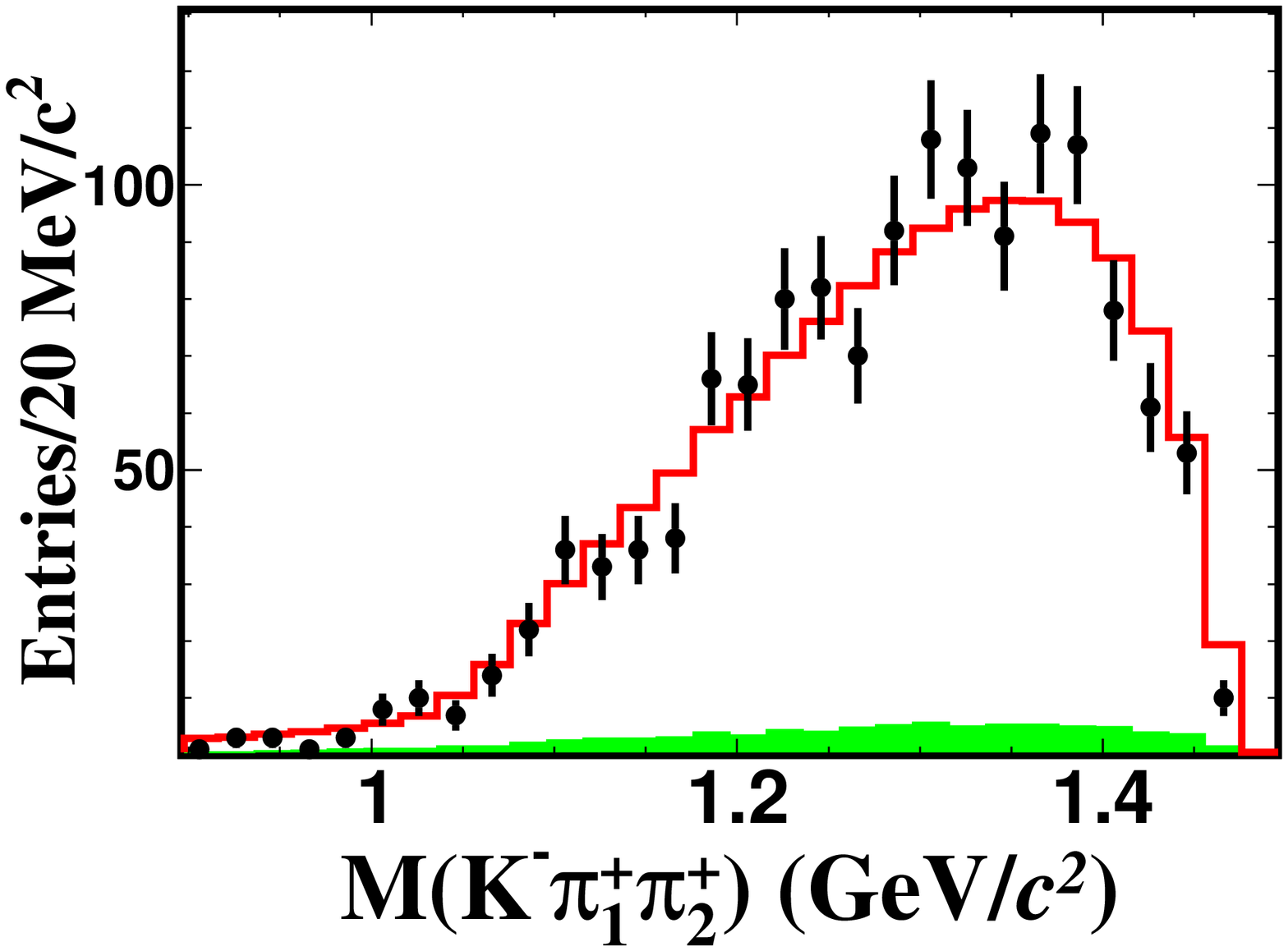}
 \put(85,60){$(i)$}
 \end{overpic}
   \caption{The projections of (a) $M_{K_S^0K^-}$, (b) $M_{K_S^0\pi^+_1}$, (c) $M_{K^-\pi^+_2}$, (d) $M_{K_S^0\pi^+_2}$, (e) $M_{K^-\pi^+_1}$, (f) $M_{K_S^0K^-\pi^+_1}$, (g) $M_{K_S^0K^-\pi^+_2}$, (h) $M_{\pi^-_1\pi^+_2}$ and (i) $M_{K^-\pi^+_1\pi^+_2}$ for the nominal amplitude fit are shown from data samples at $E_{\rm cm}$ between 4.178 and 4.226 GeV. The black points with error bars are data, the red histograms are the results of the nominal amplitude fit, the green shaded histograms are the scaled GMC combinatorial background. For the identical pions, the one giving a lower $K^0_S\pi^+$ invariant mass is denoted as $\pi^+_1$, the other is denoted as $\pi^+_2$.}
\label{pwa_plot}
\end{figure*}

In the fit, the magnitude ($\rho$) and phase ($\phi$) of $D^{+}_{s}\rightarrow K^*(892)^+\overline{K}^*(892)^0$ with angular momentum $L = 0$ between $K^*(892)^+$ and $\overline{K}^*(892)^0$ is fixed to 1 and 0, respectively, and the magnitudes and phases of the other contributions are kept floating. 
The masses and widths of all resonances are fixed to the corresponding PDG averages~\cite{PDG}. We consider possible resonant contributions from $a_0(980)$, $K^*(892)$, $K^*(1410)$, $K^*(1430)$, $K_1(1270)$, $K_1(1400)$, $\eta(1295)$, $\eta(1405)$, $\eta(1475)$, $f_1(1285)$, $f_1(1420)$, $f_1(1510)$ and $\phi(1680)$ as well as non-resonant contributions. The isospin symmetry requires the magnitude and phases  of the processes $D_{s}^{+}\rightarrow \eta(1475)\pi^{+}, \eta(1475)\rightarrow\overline{K}^{*}(892)^{0}K^{0}_{S}$ and $D_{s}^{+}\rightarrow \eta(1475)\pi^{+}, \eta(1475)\rightarrow K^{*}(892)^{+}K^{-}$ to be the same~\cite{PRL081803}. Resonant or non-resonant contributions with a significance of larger than four standard deviations are retained in the model, where the significance is calculated from the change of the log-likelihood values and the corresponding degrees of freedom. Eleven amplitude contributions are retained in the nominal fit, including a non-resonant component of $K^*(892)^+K^{-}$ with $L = 1$ between $K^*(892)^+$ and $K^{-}$. All the resonant and non-resonant contributions and their $\phi$, FFs and significances are listed in Table~\ref{pwa_result}. The magnitude and correlation matrix are provided in the supplemental material~\cite{Supplemental}. The projections for the nine invariant-mass distributions are shown in Fig.~\ref{pwa_plot}.
\begin{table*}[!hbtp]
 \centering
	\caption{The $\phi$, FFs and significances for different resonant contributions, labeled as I, II, III, $\cdots$, XIII, respectively. The first and second uncertainties are the statistical and systematic uncertainties, respectively. Here $K^*(892)^+$, $\overline{K}^*(892)^0$ and $a_0(980)^-$ denote $K^*(892)^+\rightarrow K_S^0\pi^+$, $\overline{K}^*(892)^0\rightarrow K^-\pi^+$ and $a_0(980)^-\rightarrow K_S^0K^-$, respectively, while $K(892)^*K$ indicates $\overline{K}^{*0}K_S^0$ and $K^*(892)^+K^-$. The FF of IV (IIX) term is the sum of I, II and III (VIII and IX) terms after considering the interference.
}
\renewcommand\arraystretch{1.3}
\begin{tabular}{cl r@{~$\pm$~}c@{~$\pm$~}l   r@{~$\pm$~}c@{~$\pm$~}l c}
\hline\hline
Label &\multicolumn{1}{c}{Component}   &\multicolumn{3}{c}{$\phi$} &\multicolumn{3}{c}{FF(\%)} & Significance ($\sigma$)  \\
\hline
I & $D^{+}_{s}[S]\rightarrow K^{*}(892)^{+} \overline{K}^{*}(892)^{0}$ & \multicolumn{3}{c}{0 (fixed)} & 34.3 &  3.1  &  5.2 & $>$10.0 \\  
\hline
II&$D^{+}_{s}[P]\rightarrow K^{*}(892)^{+} \overline{K}^{*}(892)^{0}$&-1.61&0.08&0.03& 7.5& 1.1& 0.1& 8.3 \\  
\hline
III&$D^{+}_{s}[D]\rightarrow K^{*}(892)^{+} \overline{K}^{*}(892)^{0}$&-0.16&0.14&0.04& 4.5& 0.8& 0.3& 8.2 \\  
\hline
IV&$D^{+}_{s}\rightarrow K^{*}(892)^{+} \overline{K}^{*}(892)^{0}$&\multicolumn{3}{c}{}&40.6& 2.9& 4.9&  \\  
\hline
V&$D_{s}^{+}\rightarrow K^{*}(892)^{+}(K^{-}\pi^{+})_{S-wave}$&1.85&0.15&0.09& 5.0& 1.2& 1.0& 6.2 \\  
\hline
VI&$D^{+}_{s}\rightarrow \overline{K}^{*}(892)^{0}(K_S^0\pi^+)_{S-wave}$&-1.57&0.12&0.13& 7.3& 1.1& 0.9& 9.1 \\  
\hline
VII&$D^{+}_{s}\rightarrow \eta(1475)\pi^+, \eta(1475)\rightarrow a_{0}(980)^{-}\pi^+$&-1.95&0.15&0.07&10.8& 2.6& 5.2& 4.4 \\  
\hline
VIII&$D_{s}^{+}\rightarrow \eta(1475)\pi^{+}, \eta(1475)\rightarrow\overline{K}^{*}(892)^{0}K^{0}_{S}$&0.05&0.15&0.11& 2.2& 0.6& 0.2& 4.5 \\  
\hline
IX&$D_{s}^{+}\rightarrow \eta(1475)\pi^{+}, \eta(1475)\rightarrow K^{*}(892)^{+}K^{-}$&0.05&0.15&0.11& 2.2& 0.6& 0.2& 4.5 \\  
\hline
IIX&$D_{s}^{+}\rightarrow \eta(1475)\pi^{+}, \eta(1475)\rightarrow K^{*}(892)K$&\multicolumn{3}{c}{}& 4.9& 1.4& 1.0&  \\  
\hline
IIIX&$D_{s}^{+}\rightarrow \eta(1475)\pi^{+}, \eta(1475)\rightarrow (K_S^0\pi^{+})_{S-wave}K^{-}$&2.30&0.11&0.07&23.6& 3.6& 7.5& 6.7 \\  
\hline
X&$D^{+}_{s}\rightarrow f_{1}(1285)\pi^+, f_{1}(1285)\rightarrow a_{0}(980)^{-}\pi^+$&-0.89&0.26&0.14& 2.2& 0.5& 0.2& 6.0 \\  
\hline
XI&\makecell[l]{$D^{+}_{s}\rightarrow (K^{*}(892)^{+}K^-)_P\pi^+$,\\$(K^{*}(892)^{+}K^-)_P\rightarrow K^{*}(892)^{+}K^-$}&-1.07&0.11&0.03&10.8& 1.9& 1.7& 9.2 \\  
\hline
\hline
\end{tabular}
 \label{pwa_result}
\end{table*}

To validate the fit performance, 300 sets of SMC samples with the same size as the data samples are generated according to the nominal fit results in this analysis. Each sample is analyzed with the same
method as for data. The pull value is given by $V_{\rm pull} = (V_{\rm fit}-V_{\rm input})/\sigma_{\rm fit}$, where $V_{\rm input}$ is the input value in the generator, $V_{\rm fit}$ and $\sigma_{\rm fit}$ is the output value and the corresponding statistical uncertainty, respectively. The resulting pull distributions are fitted with Gaussian distributions. The fitted mean value of the pull distribution for the FF of $D_{s}^{+}\rightarrow \eta(1475)\pi^{+}, \eta(1475)\rightarrow (K_S^0\pi^{+})_{S{\text-}{\rm wave}}K^{-}$ deviates from zero by more than 3.0$\sigma$, we correct its FF according to the deviation and the uncertainty of the FF.
\subsection{Systematic Uncertainties}
The systematic uncertainties for the amplitude analysis are studied in the following categories.
\begin{itemize}
  \item $K\pi$ $S$-wave model. The fixed parameters of the model are evaluated by varying the input values within $\pm 1 \sigma$ according to Ref.~\cite{PRD112012}.
	\item Lineshape of $a_0(980)$. The Flatt\'{e} parameters are shifted by $\pm$1$\sigma$ based on the values given in Ref~.\cite{PhysRevD.78.032002}.
	\item Effective barrier radius. The barrier radius are varied within $\pm$1 GeV$^{-1} \times \hbar c$ for intermediate resonances and the $D_s^+$ meson.
  \item Masses and widths of the resonances considered. The masses and widths are shifted by $\pm$1$\sigma$ based on their values from the PDG~\cite{PDG}.
  \item Background estimation. We shift the fractions of the signal in Eq.~\ref{bkg_frac} according to the uncertainty associated with the background estimation and take the largest shift as the systematic uncertainty.
  \item Experimental effects. To determine the systematic uncertainty due to tracking and PID efficiencies, we alter the fit by shifting the $\gamma_{\epsilon}$ in Eq.~\ref{pid_fun} within its uncertainty, and the change of the nominal fit result is taken as the systematic uncertainty.
  \item Neglected resonances. The intermediate processes with statistical significance less than four standard deviations are added one-by-one to the nominal contributions. For each parameter, the maximum difference with respect to the nominal fit result is taken as the corresponding systematic uncertainty.
  \item Fit uncertainties. The fitted widths from the pull distributions described in Sec.~\ref{fit results} are consistent with 1.0 within $2.0\sigma$. Therefore, the fit uncertainties are estimated properly and no systematic uncertainty is assigned from this source.
\end{itemize}

All of the systematic uncertainties of the $\phi$ and FFs are listed in Table \ref{total_sys}. The total systematic uncertainties are obtained by adding the above systematic uncertainties in quadrature.
\begin{table*}[!hbtp]
 \centering
	\caption{Summary of systematic uncertainties on the $\phi$ and FFs  from different sources, in units of the corresponding statistical uncertainties: (I) $K\pi$ $S$-wave model, (II) lineshape of $a_0(980)$, (III) effective barrier radius, (IV) masses and widths of the resonances considered, (V) background estimation, (VI) experimental effects, (VII) neglected resonances.}
\renewcommand\arraystretch{1.3}
\begin{tabular}{lccccccccc}
\hline\hline
	\multicolumn{1}{c}{\multirow{2}*{Component}}&  &\multicolumn{7}{c}{Source} &\\
			 &  &I  & II & III & IV &V &VI&VII & Total   \\
\hline
\multirow{1}*{$D^{+}_{s}[S]\rightarrow K^{*}(892)^{+} \overline{K}^{*}(892)^{0}$} & FF & 0.18 & 0.12 & 0.41 & 0.43 & 0.09&0.04&1.55 & 1.67  \\
\hline
\multirow{2}*{$D^{+}_{s}[P]\rightarrow K^{*}(892)^{+} \overline{K}^{*}(892)^{0}$} & $\phi$ & 0.02 & 0.03 & 0.03 & 0.06 & 0.01&0.00&0.43 & 0.44 \\
 & FF & 0.05 & 0.00 & 0.06 & 0.10 & 0.00&0.00&0.02 & 0.13 \\
\hline
\multirow{2}*{$D^{+}_{s}[D]\rightarrow K^{*}(892)^{+} \overline{K}^{*}(892)^{0}$} & $\phi$ & 0.02 & 0.04 & 0.00 & 0.03 & 0.01&0.03&0.28 & 0.29 \\
 & FF & 0.04 & 0.05 & 0.26 & 0.06 & 0.00&0.00&0.23 & 0.36 \\
\hline
\multirow{1}*{$D^{+}_{s}\rightarrow K^{*}(892)^{+} \overline{K}^{*}(892)^{0}$} & FF & 0.18 & 0.00 & 0.24 & 0.52 & 0.00&0.00&1.55 & 1.66  \\
\hline
\multirow{2}*{$D_{s}^{+}\rightarrow K^{*}(892)^{+}(K^{-}\pi^{+})_{S{\text-}{\rm wave}}$} & $\phi$ & 0.36 & 0.13 & 0.04 & 0.20 & 0.01&0.09&0.37 & 0.58 \\
 & FF & 0.08 & 0.05 & 0.06 & 0.02 & 0.00&0.00&0.85 & 0.86 \\
\hline
\multirow{2}*{$D^{+}_{s}\rightarrow \overline{K}^{*}(892)^{0}(K_S^0\pi^+)_{S{\text-}{\rm wave}}$} & $\phi$ & 0.48 & 0.08 & 0.06 & 0.24 & 0.01&0.03&0.87 & 1.03 \\
 & FF & 0.08 & 0.02 & 0.04 & 0.17 & 0.01&0.00&0.79 & 0.81 \\
\hline
\multirow{2}*{$D^{+}_{s}\rightarrow \eta(1475)\pi^+, \eta(1475)\rightarrow a_{0}(980)^{-}\pi^+$} & $\phi$ & 0.01 & 0.35 & 0.00 & 0.12 & 0.02&0.05&0.26 & 0.45 \\
 & FF & 0.05 & 1.96 & 0.12 & 0.22 & 0.05&0.02&0.08 & 1.98 \\
\hline
\multirow{2}*{$D_{s}^{+}\rightarrow \eta(1475)\pi^{+}, \eta(1475)\rightarrow\overline{K}^{*}(892)^{0}K^{0}_{S}$} & $\phi$ & 0.01 & 0.13 & 0.09 & 0.70 & 0.00&0.03&0.12 & 0.73 \\
 & FF & 0.02 & 0.02 & 0.02 & 0.04 & 0.00&0.00&0.29 & 0.30 \\
\hline
\multirow{2}*{$D_{s}^{+}\rightarrow \eta(1475)\pi^{+}, \eta(1475)\rightarrow K^{*}(892)^{+}K^{-}$} & $\phi$ & 0.01 & 0.13 & 0.09 & 0.70 & 0.00&0.03&0.12 & 0.73 \\
 & FF & 0.02 & 0.02 & 0.02 & 0.04 & 0.00&0.00&0.31 & 0.31 \\
\hline
	\multirow{1}*{$D_{s}^{+}\rightarrow \eta(1475)\pi^{+}, \eta(1475)\rightarrow K^{*}(892)K$} & FF & 0.04 & 0.05 & 0.05 & 0.09 & 0.01&0.00&0.66 & 0.68  \\
\hline
\multirow{2}*{$D_{s}^{+}\rightarrow \eta(1475)\pi^{+}, \eta(1475)\rightarrow (K_S^0\pi^{+})_{S{\text-}{\rm wave}}K^{-}$} & $\phi$ & 0.48 & 0.08 & 0.01 & 0.40 & 0.01&0.07&0.26 & 0.68 \\
 & FF & 0.25 & 0.62 & 0.23 & 0.80 & 0.00&0.00&0.46 & 1.16 \\
\hline
\multirow{2}*{$D^{+}_{s}\rightarrow f_{1}(1285)\pi^+, f_{1}(1285)\rightarrow a_{0}(980)^{-}\pi^+$} & $\phi$ & 0.03 & 0.05 & 0.02 & 0.18 & 0.00&0.02&0.48 & 0.52 \\
 & FF & 0.01 & 0.45 & 0.01 & 0.08 & 0.00&0.00&0.12 & 0.48 \\
\hline
	\multirow{2}*{\makecell[l]{$D^{+}_{s}\rightarrow (K^{*}(892)^{+}K^-)_P\pi^+$,\\$(K^{*}(892)^{+}K^-)_P\rightarrow K^{*}(892)^{+}K^-$}} & $\phi$ & 0.01 & 0.03 & 0.06 & 0.09 & 0.00&0.02&0.21 & 0.24 \\
 & FF & 0.05 & 0.19 & 0.18 & 0.03 & 0.06&0.02&0.87 & 0.91 \\
\hline
\hline
\end{tabular}
 \label{total_sys}
\end{table*}
\section{BRANCHING-FRACTION MEASUREMENT}
\subsection{Yields and Efficiencies}\label{bf_signal}
The selection criteria of the tagged $D_s^-$ and signal $D_s^+$ candidates are the same as in Sec.~\ref{tag_selection}, except for the following requirements: (I) the requirement of the secondary vertex fit for $K^0_S$ from the tag modes is removed, while that for the signal is retained; (II) a further requirement of $p_{\pi^{\pm}/\pi^0}>$ 0.1~GeV/$c$ is added to remove the soft $\pi^{\pm}/\pi^0$ directly from $D^{*\pm}/D^{*0}$ decays; (III) the tagged $D_{s}^-$ candidates are reconstructed by looping over all their daughter tracks to form different combinations. If there are multiple candidates from the same event, the one with $M_{\rm rec}$ closest to the $D_s^{*\pm}$ mass is accepted; (IV) at least one of the $D_s^+/D_s^-$ candidates must satisfy $M_{\rm rec}>2.10$~GeV/$c^2$;
(V) the combination with average mass $\overline{M} = [M_{\rm tag}+M_{\rm sig}]/2$ closest to the nominal mass of $D_s^+$ \cite{PDG} is chosen among the multiple candidates. 

\begin{figure*}[!hbtp]
\includegraphics[scale=0.8]{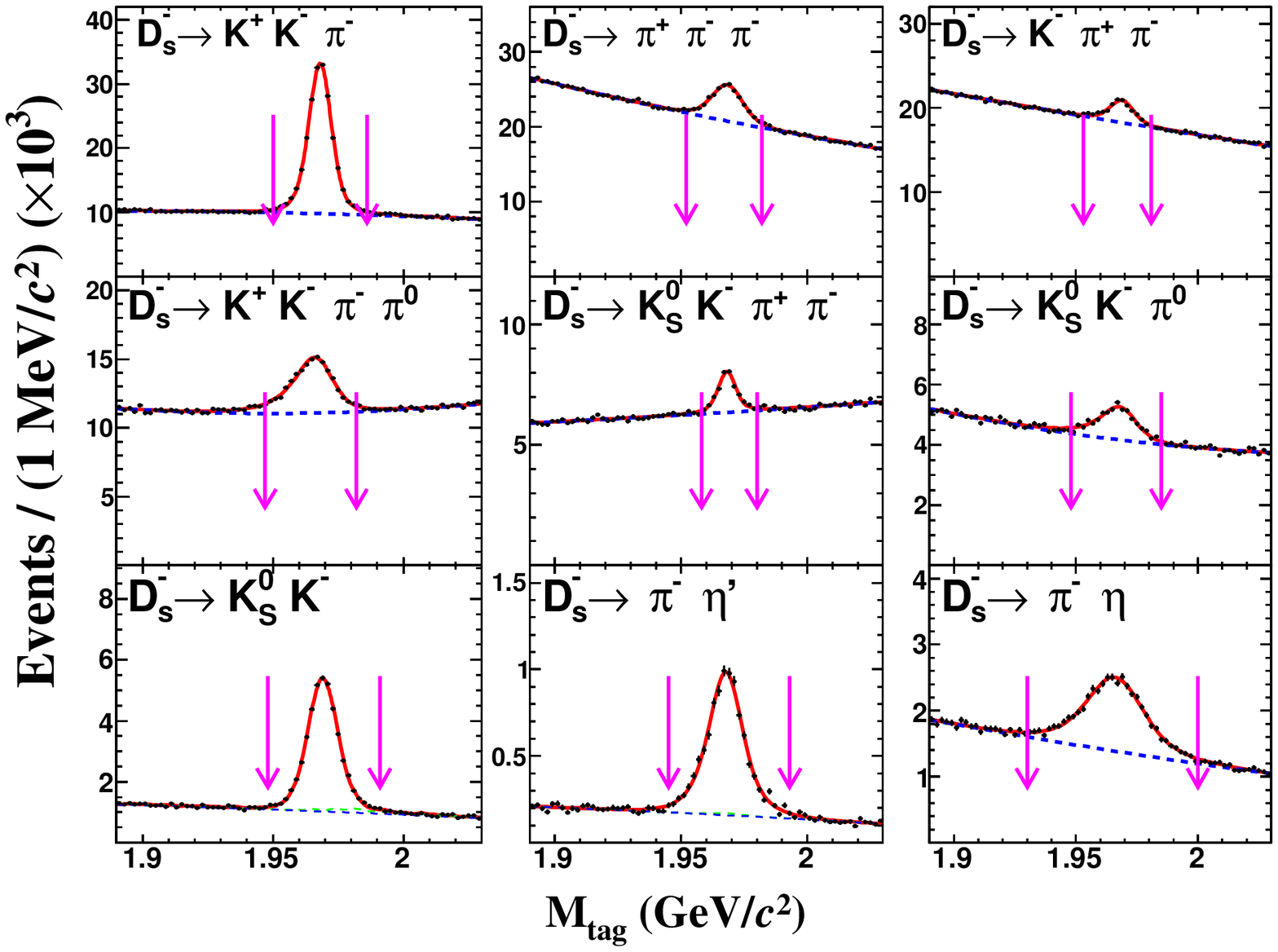}
\caption{Best fit results to the $M_{\rm tag}$ distributions of the ST candidates from the data sample taken at $E_{\rm cm}$ = 4.178 GeV. The points with error bars are data. The red solid curves are the fit results. The blue dotted curves are the fitted background shapes. The pair of pink arrows indicates the chosen signal regions. The green dotted curve in the $D_s^-\rightarrow K_S^0K^-$ ($D_s^-\rightarrow \pi^-\eta^\prime$) mode is the $D_s^-\rightarrow K_S^0\pi^-$ ($D_s^-\rightarrow \pi^+\pi^-\pi^-\eta$) component.}
\label{tag_yield18}
\end{figure*}

\begin{figure}[htb]
 \centering
 \includegraphics[scale=0.35]{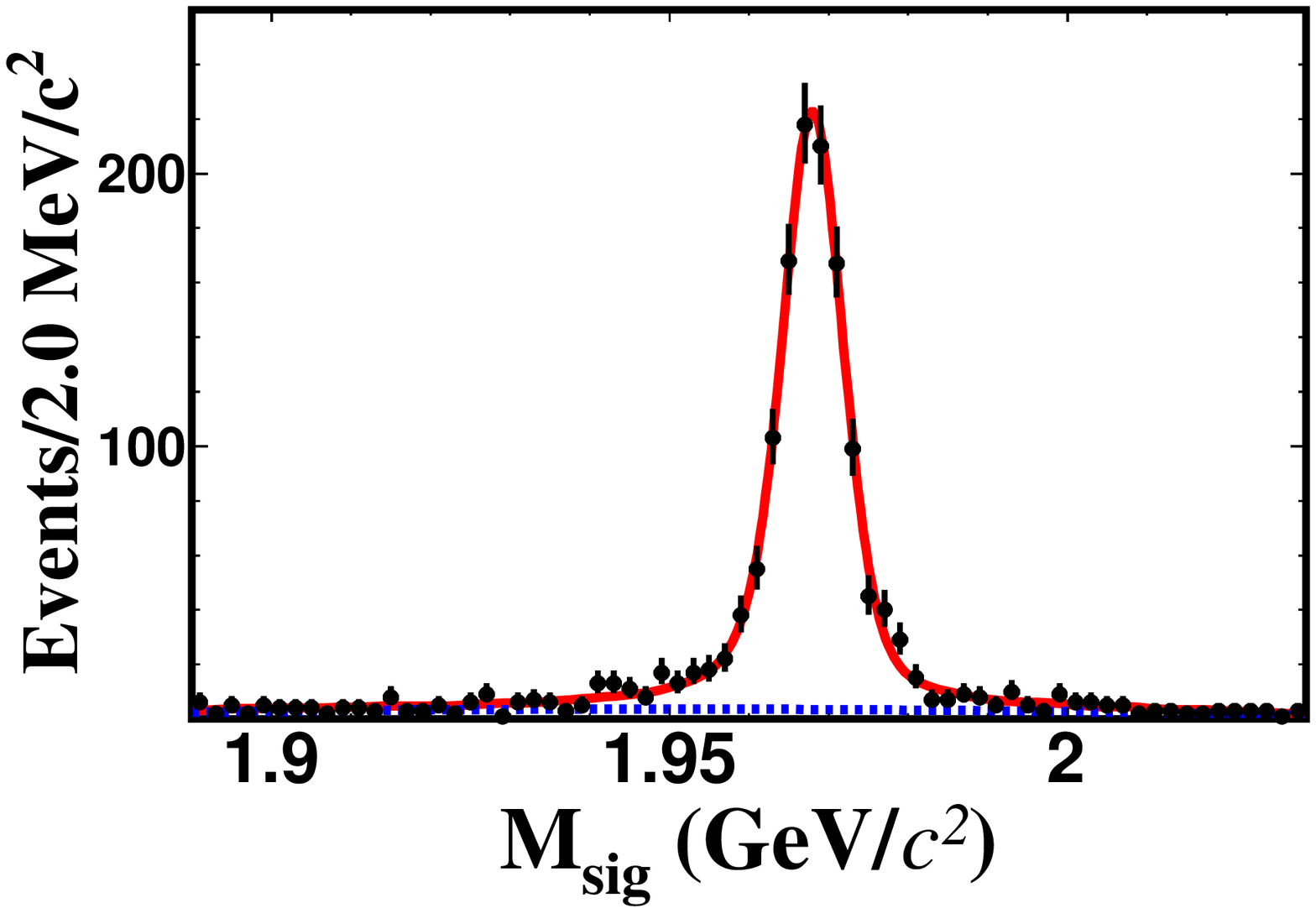}\\
	 \caption{Best fit result to the $M_{\rm sig}$ distributions of the DT candidates from data samples at $E_{\rm cm}$ between 4.178 and 4.226~GeV. The points with error bars are data. The red solid curve is the fit result. The blue dotted curve is the fitted background shape.}
 \label{62Data_sigma}
\end{figure}

The ST yields ($N_{\rm ST}$) and DT yields ($N_{\rm DT}$) in data are determined by fitting the $M_{\rm tag}$ distributions from different tag modes and $M_{\rm sig}$ distributions, respectively. In each fit, the signal shape is modeled using the simulated shape convolved with a Gaussian function, whose resolution and mean are free parameters, and the background is described with a second-order Chebychev polynomial. These fits give a total ST yield of $N_{\rm ST}$ = 550496 $\pm$ 2411. The $M_{\rm tag}$ distributions at $E_{\rm cm}$ = 4.178 GeV are shown in Fig.~\ref{tag_yield18} as an example.  The total DT signal yield, $N_{\rm DT}^{\rm tot}$, is determined to be  $1332 \pm 42$, as shown in Fig.~\ref{62Data_sigma}.
The fits to the $M_{\rm sig}$ distribution for GMC are performed to estimate the corresponding ST efficiencies ($\epsilon_{\rm ST}$). The DT efficiencies ($\epsilon_{\rm DT}$) are determined by GMC, in which our amplitude analysis model is taken for the generation of the signal mode.
\subsection{Tagging Technique and Branching Fraction}
The branching fraction for the signal mode is given by
\begin{eqnarray}
\begin{split}
\mathcal{B}_{\rm sig}=\frac{N^{\rm tot}_{\rm DT}}{\sum\limits_{i}\sum\limits_{j}N^{ij}_{\rm ST}\cdot\epsilon^{ij}_{\rm DT}/\epsilon^{ij}_{\rm ST}},
\end{split}
\label{eq3}
\end{eqnarray}
where the indices $i$ and $j$ denote the $i^{\rm th}$ tag mode and the $j^{\rm th}$ center-of-mass energy point. The $N^{ij}_{\rm ST}$ and $\epsilon^{ij}_{\rm ST(\rm DT)}$ are the number of the ST candidates and the corresponding ST (DT) detection efficiency.

Using Eq.~\ref{eq3} and the PDG value of the $\mathcal{B}(K^{0}_{S}\rightarrow\pi^{+}\pi^{-})$ = (69.20$\pm$0.05)\%~\cite{PDG}, the absolute branching fraction can be obtained 
\begin{eqnarray}
\begin{split}
\mathcal{B}(D^{+}_{s}\rightarrow K^{0}_{S}K^{-}\pi^{+}\pi^{+}) = (1.46\pm0.05)\%,
\end{split}
\label{eq4}
\end{eqnarray}where the uncertainty is statistical.

\subsection{SYSTEMATIC UNCERTAINTIES}
The systematic uncertainties for the branching fraction measurement are studied in the following categories.
\begin{itemize}
 \item $K^{\pm}$ and $\pi^{\pm}$ tracking (PID) efficiencies. The tracking (PID) efficiencies are studied using  samples of $e^+e^- \rightarrow K^+K^-\pi^+\pi^-$ ($e^+e^- \rightarrow K^+K^-K^+K^-$, $K^+K^-\pi^+\pi^-(\pi^0)$ and $\pi^+\pi^-\pi^+\pi^-(\pi^0)$) events. The systematic uncertainties for $K^{\pm}$ and $\pi^{\pm}$ due to tracking (PID) are estimated to be 0.8\% and 0.3\% (0.8\% and 0.5\%), respectively.
 \item $K^0_{S}$ reconstruction efficiency. The uncertainty for the $K^0_{S}$ reconstruction efficiency is assigned as 1.5\% per $K^0_{S}$, obtained using control samples of $J/\psi\to$ $K^0_S K^\pm\pi^\mp$ and $\phi K^0_S K^\pm\pi^\mp$ decays.
 \item Fit to the DT $M_{\rm sig}$ distribution. The uncertainty associated with the modeling of the DT $M_{\rm sig}$ distribution is studied with alternative models for signal and background. The uncertainties are estimated by comparing with the fit results obtained using the signal and background shapes directly from the simulated samples. 
 \item Fit to the ST $M_{\rm tag}$ distribution. We change the background shape from the second-order Chebychev polynomial to a third-order Chebychev polynomial, causing a 0.18\% relative change of the branching fraction. The systematic uncertainty due to the modeling of the signal distribution is determined to be 0.16\% by performing an alternative fit using the shape directly obtained from the simulated sample. The quadratic sum of these terms, 0.24\%, is assigned as the systematic uncertainty.
 \item Measurement method. The possible bias due to the measurement method is estimated to be 0.3\% by comparing the measured branching fraction in the SMC, using the same method as in data analysis, to the value input in the SMC generation.
 \item Statistics of simulated events. The uncertainty associated with the limited statistics of GMC for the detection efficiency is 0.3\%.
 \item Amplitude analysis model. The uncertainty from the amplitude analysis model is 0.6\%, estimated from the efficiency difference obtained by varying the fitted parameters $c_n$ in Eq.~\ref{rho_phi} according to the error matrix. 
\end{itemize}

All the systematic uncertainties of the branching fraction measurement are listed in Table~\ref{bfsys}. When added in quadrature they sum to a relative uncertainty of $3.3\%$, which is the same as the statistical uncertainty on the measurement.
\begin{table}[!hbtp]
 \renewcommand\arraystretch{1.25}
 \centering
 \caption{ Systematic uncertainties in the branching-fraction measurement.}
 \begin{tabular}{c|c}
  \hline \hline
  Source & Uncertainty (\%) \\
  \hline
  Tracking efficiency& 1.4 \\
   \hline
  PID efficiency& 1.8 \\
   \hline
  $K^0_S$ reconstruction efficiency & 1.5 \\
   \hline
  DT $M_{\rm sig}$ fit & 1.7 \\
   \hline
   ST $M_{\rm tag}$ fit & 0.2 \\
   \hline
Measurement method & 0.3 \\
   \hline
   Statistics of simulated events & 0.3 \\
   \hline
  Amplitude analysis model & 0.6 \\
   \hline
  $\mathcal{B}(K^{0}_{S}\rightarrow\pi^{+}\pi^{-})$~\cite{PDG} & 0.1 \\
  \hline
  Total  & 3.3 \\
  \hline \hline
 \end{tabular}
 \label{bfsys}
\end{table}
\section{CONCLUSION}
Using 6.32 fb$^{-1}$ of $e^+e^-$ collision data collected by the BESIII detector with center-of-mass energies between 4.178 and 4.226~GeV, we report the first amplitude analysis of the $D^{+}_{s}\rightarrow K^{0}_{S}K^{-}\pi^{+}\pi^{+}$ decays and an improved measurement of the  $D^{+}_{s}\rightarrow K^{0}_{S}K^{-}\pi^{+}\pi^{+}$ branching fraction. 
The model indicates that the quasi-two-body decay $D^{+}_{s}\rightarrow K^*(892)^+\overline{K}^*(892)^0$ is dominant, with a fit fraction of ($40.6\pm2.9_{\rm stat}\pm4.9_{\rm sys})$\%.  
In addition, there are significant contributions from $f_1(1285)$, $\eta(1475)$ and $(K^{*}(892)^{+}K^-)_P$ in the mass spectrum of $K^{0}_{S}K^{-}\pi^{+}$. The $\eta(1475)$ meson decays to both $K^*K$ and $a_0(980)\pi$ final states, while the $f_1(1285)$ meson decays only to $a_0(980)\pi$. 
The absolute branching fraction of the $D^+_s\to K^0_SK^-\pi^+\pi^+$ decay is determined to be $(1.46\pm0.05_{\rm stat}\pm0.05_{\rm sys}$)\%, and the branching fractions for different components are listed in Table~\ref{compare}.
The branching fraction of the quasi-two-body  decay $D^{+}_{s}\rightarrow K^*(892)^+ \overline{K}^*(892)^0$ is calculated to be $(5.34\pm0.39_{\rm stat}\pm0.64_{\rm sys})$\%. 
Our measurements are consistent with the current world averages  \cite{PDG} but much more precise. 
 \begin{table*}[!hbtp]
 \centering
	 \caption{ The branching fractions measured in this analysis and from PDG \cite{PDG}. The $K^*(892)^+$, $\overline{K}^*(892)^0$ and $a_0(980)^-$ denote $K^*(892)^+\rightarrow K_S^0\pi^+$, $\overline{K}^*(892)^0\rightarrow K^-\pi^+$ and $a_0(980)^-\rightarrow K_S^0K^-$, respectively.} 
\renewcommand\arraystretch{1.3}
\begin{tabular}{lcc}
\hline\hline
\multicolumn{1}{c}{\multirow{2}*{Process}} & \multicolumn{2}{c}{BF(10$^{-3}$) }   \\ \cline{2-3}
  & This analysis & PDG \\ 
\hline
$D^{+}_{s}[S]\rightarrow K^{*}(892)^{+} \overline{K}^{*}(892)^{0}$ & 5.01 $\pm$ 0.49 $\pm$ 0.78 &  \\  
\hline
$D^{+}_{s}[P]\rightarrow K^{*}(892)^{+} \overline{K}^{*}(892)^{0}$ & 1.10 $\pm$ 0.16 $\pm$ 0.10 &  \\  
\hline
$D^{+}_{s}[D]\rightarrow K^{*}(892)^{+} \overline{K}^{*}(892)^{0}$ & 0.65 $\pm$ 0.12 $\pm$ 0.10 &  \\  
\hline
$D^{+}_{s}\rightarrow K^{*}(892)^{+} \overline{K}^{*}(892)^{0}$ & 5.93 $\pm$ 0.47 $\pm$ 0.74 & 7.98 $\pm$ 2.88  \\  
\hline
$D_{s}^{+}\rightarrow K^{*}(892)^{+}(K^{-}\pi^{+})_{S-{\rm wave}}$ & 0.73 $\pm$ 0.17 $\pm$ 0.15 &  \\  
\hline
$D^{+}_{s}\rightarrow \overline{K}^{*}(892)^{0}(K_S^0\pi^+)_{S-{\rm wave}}$ & 1.06 $\pm$ 0.16 $\pm$ 0.13 &  \\  
\hline
$D^{+}_{s}\rightarrow \eta(1475)\pi^+, \eta(1475)\rightarrow a_{0}(980)^{-}\pi^+$ & 1.57 $\pm$ 0.39 $\pm$ 0.76 &  \\  
\hline
$D_{s}^{+}\rightarrow \eta(1475)\pi^{+}, \eta(1475)\rightarrow\overline{K}^{*}(892)^{0}K^{0}_{S}$ & 0.32 $\pm$ 0.10 $\pm$ 0.10 &  \\  
\hline
$D_{s}^{+}\rightarrow \eta(1475)\pi^{+}, \eta(1475)\rightarrow K^{*}(892)^{+}K^{-}$ & 0.32 $\pm$ 0.10 $\pm$ 0.10 &  \\  
\hline
$D_{s}^{+}\rightarrow \eta(1475)\pi^{+}, \eta(1475)\rightarrow K^{*}(892)K$ & 0.72 $\pm$ 0.21 $\pm$ 0.14 &  \\  
\hline
$D_{s}^{+}\rightarrow \eta(1475)\pi^{+}, \eta(1475)\rightarrow (K_S^0\pi^{+})_{S-{\rm wave}}K^{-}$ & 3.44 $\pm$ 0.54 $\pm$ 1.10 &  \\  
\hline
$D^{+}_{s}\rightarrow f_{1}(1285)\pi^+, f_{1}(1285)\rightarrow a_{0}(980)^{-}\pi^+$ & 0.33 $\pm$ 0.08 $\pm$ 0.10 &  \\  
\hline
	\makecell[l]{$D^{+}_{s}\rightarrow (K^{*}(892)^{+}K^-)_P\pi^+$,\\$(K^{*}(892)^{+}K^-)_P\rightarrow K^{*}(892)^{+}K^-$} & 1.58 $\pm$ 0.28 $\pm$ 0.26 &  \\  
\hline
$D^{+}_{s}\rightarrow K_S^0K^{-}\pi^+\pi^+$ &  14.60 $\pm$ 0.46 $\pm$ 0.48 & 16.50 $\pm$ 1.00 \\ 
\hline
\hline
\end{tabular}
 \label{compare}
\end{table*}
\section*{ACKNOWLEDGMENTS}
The BESIII collaboration thanks the staff of BEPCII and the IHEP computing center for their strong support. This work is supported in part by National Key Research and Development Program of China under Contracts Nos. 2020YFA0406400, 2020YFA0406300; National Natural Science Foundation of China (NSFC) under Contracts Nos. 11625523, 11635010, 11735014, 11822506, 11835012, 11935015, 11935016, 11935018, 11961141012; the Chinese Academy of Sciences (CAS) Large-Scale Scientific Facility Program; Joint Large-Scale Scientific Facility Funds of the NSFC and CAS under Contracts Nos. U1732263, U1832107, U1832207, U2032104; CAS Key Research Program of Frontier Sciences under Contracts Nos. QYZDJ-SSW-SLH003, QYZDJ-SSW-SLH040; 100 Talents Program of CAS; INPAC and Shanghai Key Laboratory for Particle Physics and Cosmology; ERC under Contract No. 758462; European Union Horizon 2020 research and innovation programme under Contract No. Marie Sklodowska-Curie grant agreement No 894790; German Research Foundation DFG under Contracts Nos. 443159800, Collaborative Research Center CRC 1044, FOR 2359, FOR 2359, GRK 214; Istituto Nazionale di Fisica Nucleare, Italy; Ministry of Development of Turkey under Contract No. DPT2006K-120470; National Science and Technology fund; Olle Engkvist Foundation under Contract No. 200-0605; STFC (United Kingdom); The Knut and Alice Wallenberg Foundation (Sweden) under Contract No. 2016.0157; The Royal Society, UK under Contracts Nos. DH140054, DH160214; The Swedish Research Council; U. S. Department of Energy under Contracts Nos. DE-FG02-05ER41374, DE-SC-0012069.

\end{document}


\vspace{3.0cm}
\centerline{ \bf \Large Supplemental  material}
\vspace{1.0cm}
\centerline{ \bf \boldmath BESIII Collaboration}

\begin{table}[htbp]
	\caption{The magnitude ($\rho$) for different resonant contributions, uncertainties are statistical only.}
  \centering
	\renewcommand\arraystretch{1.2}
  \begin{tabular}{lc}
  \hline \hline
		 \multicolumn{1}{c}{Component} & $\rho$ \\\hline
		 $D^{+}_{s}[P]\rightarrow K^{*}(892)^{+} \overline{K}^{*}(892)^{0}$    &$0.36\pm0.03$\\
		 $D^{+}_{s}[D]\rightarrow K^{*}(892)^{+} \overline{K}^{*}(892)^{0}$    &$0.47\pm0.04$\\
		\makecell[l]{ $D_{s}^{+}\rightarrow \eta(1475)\pi^{+}, \eta(1475)\rightarrow\overline{K}^{*}(892)^{0}K^{0}_{S}$ \\ and  $\eta(1475)\rightarrow\overline{K}^{*}(892)^{0}K^{0}_{S}$}    &$1.03\pm0.16$\\
   	$D^{+}_{s}\rightarrow \overline{K}^{*}(892)^{0}(K_S^0\pi^+)_{S{\text-}{\rm wave}}$   &$1.89\pm0.25$\\
	  $D_{s}^{+}\rightarrow \eta(1475)\pi^{+}, \eta(1475)\rightarrow (K_S^0\pi^{+})_{S{\text-}{\rm wave}}K^{-}$    &$4.32\pm0.40$\\
		$D^{+}_{s}\rightarrow \eta(1475)\pi^+, \eta(1475)\rightarrow a_{0}(980)^{-}\pi^+$    &$1.36\pm0.18$\\
		$D^{+}_{s}\rightarrow f_{1}(1285)\pi^+, f_{1}(1285)\rightarrow a_{0}(980)^{-}\pi^+$    &$1.79\pm0.23$\\
		$D^{+}_{s}\rightarrow \overline{K}^{*}(892)^{0}(K_S^0\pi^+)_{S{\text-}{\rm wave}}$    &$2.19\pm0.20$\\
		\makecell[l]{$D^{+}_{s}\rightarrow (K^{*}(892)^{+}K^-)_P\pi^+$, \\ $(K^{*}(892)^{+}K^-)_P\rightarrow K^{*}(892)^{+}K^-$}    &$4.32\pm0.52$\\
        \hline
      \hline
  \end{tabular}
  \label{}
\end{table}

\begin{table*}[!hbtp]
	\centering
		 \caption{Correlation matrix. (I)$D^{+}_{s}[P]\rightarrow K^{*}(892)^{+} \overline{K}^{*}(892)^{0}$, (II)$D^{+}_{s}[D]\rightarrow K^{*}(892)^{+} \overline{K}^{*}(892)^{0}$, (III)$D_{s}^{+}\rightarrow \eta(1475)\pi^{+}, \eta(1475)\rightarrow\overline{K}^{*}(892)^{0}K^{0}_{S}$ and $\eta(1475)\rightarrow\overline{K}^{*}(892)^{0}K^{0}_{S}$, (IV)$D^{+}_{s}\rightarrow \overline{K}^{*}(892)^{0}(K_S^0\pi^+)_{S{\text-}{\rm wave}}$, (V)$D_{s}^{+}\rightarrow \eta(1475)\pi^{+}, \eta(1475)\rightarrow (K_S^0\pi^{+})_{S{\text-}{\rm wave}}K^{-}$, (VI)$D^{+}_{s}\rightarrow \eta(1475)\pi^+, \eta(1475)\rightarrow a_{0}(980)^{-}\pi^+$, (VII)$D^{+}_{s}\rightarrow f_{1}(1285)\pi^+, f_{1}(1285)\rightarrow a_{0}(980)^{-}\pi^+$, (VIII)$D^{+}_{s}\rightarrow \overline{K}^{*}(892)^{0}(K_S^0\pi^+)_{S{\text-}{\rm wave}}$, (IX)$D^{+}_{s}\rightarrow (K^{*}(892)^{+}K^-)_P\pi^+, (K^{*}(892)^{+}K^-)_P\rightarrow K^{*}(892)^{+}K^-$.}
	\renewcommand\arraystretch{1.1}
	\begin{tabular}{cc|rr|rr|rr|rr|rr|rr|rr|rr|rr}
		\hline
		\hline
		 &		& \multicolumn{2}{c|}{I} & \multicolumn{2}{c|}{II} & \multicolumn{2}{c|}{III} & \multicolumn{2}{c|}{IV} &	\multicolumn{2}{c|}{V} & \multicolumn{2}{c|}{VI} & \multicolumn{2}{c|}{VII} & \multicolumn{2}{c|}{VIII} &\multicolumn{2}{c}{IX}   \\
			&	& \multicolumn{1}{c}{$\phi$} & \multicolumn{1}{c|}{$\rho$}       & \multicolumn{1}{c}{$\phi$} & \multicolumn{1}{c|}{$\rho$}        & \multicolumn{1}{c}{$\phi$} & \multicolumn{1}{c|}{$\rho$}         & \multicolumn{1}{c}{$\phi$} & \multicolumn{1}{c|}{$\rho$}   & \multicolumn{1}{c}{$\phi$} &\multicolumn{1}{c|}{$\rho$}       & \multicolumn{1}{c}{$\phi$} & \multicolumn{1}{c|}{$\rho$}        & \multicolumn{1}{c}{$\phi$} & \multicolumn{1}{c|}{$\rho$}   & \multicolumn{1}{c}{$\phi$} & \multicolumn{1}{c|}{$\rho$}& \multicolumn{1}{c}{$\phi$}& \multicolumn{1}{c}{$\rho$}  \\ \hline

		\multirow{2}*{I}& $\phi$   &  1.00 & 0.15&  0.34 &-0.16 &   -0.03 & 0.17&  0.02 & 0.12 &    0.04& 0.01 &   0.04 & 0.04 &   0.14 & 0.12 &    0.33 & 0.03 &    0.21 &   0.10\\
		& $\rho$ & &1.00&-0.02& 0.05& 0.01& 0.31& 0.17&0.32 & 0.06&0.33& 0.03& 0.21& 0.03&0.22& 0.05&0.31 &-0.15& 0.47\\\hline
		\multirow{2}*{II}& $\phi$  &&&  1.00 &-0.31 &   -0.24 & 0.11& -0.08 & 0.17 &   -0.24&-0.17 &  -0.23 &-0.02 &   0.19 & 0.21 &    0.32 &-0.12 &    0.02 &   0.03\\
		& $\rho$ &&&& 1.00& 0.22&-0.15& 0.40&0.06 & 0.19&0.09& 0.19&-0.02& 0.04&0.03&-0.10&0.03 &-0.13& 0.21\\\hline
		\multirow{2}*{III}& $\phi$ &&&& &   1.00 &-0.24&  0.33 & 0.11 &    0.47&-0.21 &   0.44 &-0.37 &   0.14 &-0.18 &    0.27 & 0.09 &    0.12 &  -0.13\\
		& $\rho$	& &&&&& 1.00&-0.28&0.60 &-0.02&0.58& 0.13& 0.50&-0.21&0.14&-0.10&0.59 &-0.30& 0.30\\\hline
		\multirow{2}*{IV}& $\phi$	 &&&&& &&  1.00 & 0.11 &    0.38&-0.05 &   0.28 &-0.17 &   0.30 & 0.10 &    0.15 & 0.05 &   -0.13 &   0.50\\
		& $\rho$	& && & & & & &1.00 &-0.06&0.30& 0.07& 0.28& 0.01&0.16& 0.13&0.46 &-0.53& 0.34\\\hline
		\multirow{2}*{V}& $\phi$ &   & &  &  &   &&  & &    1.00& 0.07 &   0.80 &-0.23 &   0.04 &-0.09 &    0.07 &-0.04 &    0.19 &   0.15\\
		& $\rho$	& &&& && && & &1.00& 0.18& 0.81&-0.22&0.15&-0.13&0.41 &-0.23& 0.40\\\hline
		\multirow{2}*{VI}& $\phi$ &   & &  &  &    & &   &  &    &  &   1.00 &-0.07 &   0.03 &-0.07 &    0.11 & 0.04 &    0.11 &   0.11\\
		& $\rho$	& &&&&& && &&&& 1.00&-0.13&0.15&-0.12&0.30 &-0.19& 0.23\\\hline
		\multirow{2}*{VII}& $\phi$ &   & &   &  &     &&   &  &    & &    & &   1.00 & 0.19 &    0.32 &-0.11 &    0.07 &   0.05\\
		& $\rho$	& && & && & & &&&& & &1.00& 0.15&0.05 &-0.15& 0.23\\\hline
		\multirow{2}*{VIII}& $\phi$&   & &   & &     &&   &  &    & &    & &    &  &    1.00 &-0.13 &    0.14 &   0.01\\
		& $\rho$	& &&& & & & & &&& & &&&&1.00 &-0.29& 0.38\\\hline
		\multirow{2}*{IX}& $\phi$	 &   &&   & &     &&  & &    & &    & &    & &     & &    1.00 &  -0.37\\
		& $\rho$	& && & && & & & && & & && & && 1.00\\\hline\hline
	\end{tabular}
			\label{total_sys}
\end{table*}
